\documentclass[11pt]{article}
\usepackage{amsmath, latexsym,ifthen, amsfonts,amsbsy}
\usepackage{setspace,subfig}
\usepackage{natbib}
\usepackage{relsize}
\usepackage{threeparttable}
\RequirePackage[colorlinks,citecolor=blue,urlcolor=blue]{hyperref}

\renewcommand{\harvardurl}[1]{\textbf{URL:} \url{#1}}
\usepackage{booktabs}
\usepackage{pgf}
\usepackage{setspace}
\usepackage{times}
\usepackage{lipsum}
\usepackage{subfig}
\usepackage{tabularx}
   
\usepackage{amsmath,amsthm,amssymb,ifthen}
\usepackage{graphicx}
\usepackage[mathscr]{eucal}

\usepackage{enumerate}
\usepackage{etex}
\usepackage{booktabs}
\usepackage{amsfonts}
\usepackage[latin1]{inputenc}
\usepackage{soul,color}
\usepackage{chngcntr}
\usepackage{amsthm}
\usepackage{verbatim}
\usepackage{changepage}
\usepackage{amssymb}
\usepackage{adjustbox}

\usepackage{caption} 
\usepackage{xfrac}
\usepackage{graphicx}

\newtheoremstyle{note}
{8pt}
{8pt}
{}
{}
{\bfseries}
{:}
{.5em}
{}

\theoremstyle{note}
\newtheorem{theorem}{Theorem}

\newtheorem{remark}{Remark}
\usepackage[T1]{fontenc}
\newtheorem{assumption}{Assumption}
\newtheorem{definition}{Definition}
\newtheorem{example}{Example}

\newtheorem{proposition}{Proposition}
\allowdisplaybreaks

\usepackage{multirow}
\usepackage[left=1in,right=1in,top=1in,bottom=1in]{geometry}
\usepackage{algorithm}
\date{}
\usepackage{soulpos}
\soulregister\cite7
\soulregister\citep7
\soulregister\ref7
\soulregister\emph7
\soulregister\eqref7
\soulregister\pageref7
\definecolor{mygreen}{RGB}{144,241,47}

\def\T{ {\mathrm{\scriptscriptstyle T}} }

\makeatletter
\newcommand*{\centernot}{%
  \mathpalette\@centernot
}
\def\@centernot#1#2{%
  \mathrel{%
    \rlap{%
      \settowidth\dimen@{$\m@th#1{#2}$}%
      \kern.5\dimen@
      \settowidth\dimen@{$\m@th#1=$}%
      \kern-.5\dimen@
      $\m@th#1\not$%
    }%
    {#2}%
  }%
}
\makeatother
\def\bSig\mathbf{\Sigma}

\usepackage[figuresright]{rotating}

\usepackage{authblk}
\author[1]{Baoluo Sun}
\author[2]{Zhonghua Liu}
\author[3]{Eric Tchetgen Tchetgen}
\affil[1]{Department of Statistics and Data Science, National University of Singapore}
\affil[2]{Department of Statistics and Actuarial Science, 
	University of Hong Kong}
\affil[3]{Department of Statistics and Data Science, The Wharton School, University of Pennsylvania}
	{\raise0.5ex\hbox{$#1$}\! \left/ \! \lower0.5ex\hbox{$#2$}\right.}
	
\begin{document}

\title{\bf Semiparametric Efficient G-estimation with Invalid Instrumental Variables}

	\clearpage \maketitle
	

\begin{abstract}	
		
The instrumental variable method is widely used in the health and social sciences for identification and estimation of causal effects in the presence of potential unmeasured confounding. In order to improve efficiency, multiple instruments
are routinely used, leading to concerns about bias due to possible
violation of the instrumental variable assumptions. 
To address this concern, we introduce a
new class of  g-estimators that are guaranteed to remain consistent and asymptotically normal for
the causal effect of interest
provided that  a set of at least
$\gamma$ out of $K$ candidate instruments are valid, for
$\gamma\leq K$ set by the analyst {\it ex ante}, without necessarily knowing the identity of the valid and invalid instruments. 
We provide formal semiparametric efficiency theory supporting our results.
Both  simulation studies and  applications to the UK Biobank data demonstrate the superior empirical performance of our estimators compared to  competing methods.
\end{abstract}


\section{Introduction}
\label{sec: intro}

Mendelian randomization is an instrumental variable (IV) approach to causal inference with growing
popularity in epidemiological studies. In Mendelian randomization studies, one aims to establish a causal
relationship between a given exposure and an outcome of interest in the
presence of possible unmeasured confounding, by leveraging one or more genetic
markers defining the IV \citep{Davey-Smith:2003aa,Lawlor:2008aa}.
 In order to be a valid IV, a genetic marker must satisfy the following
key conditions: (a) It must be associated with the exposure; (b) It must be independent of any unmeasured confounder of the
exposure-outcome relationship; and (c) It cannot have a direct effect on the outcome variable not fully mediated by the exposure in view.  Possible violation or near violation of assumption (a) known as the
weak instrumental variable problem  poses an important challenge in Mendelian randomization  as
individual genetic effects on complex traits can be weak. Violation of assumption (b) can also
occur due to linkage disequilibrium or population stratification \citep{Lawlor:2008aa}.
Assumption (c) also known as the exclusion restriction is rarely credible in
the Mendelian randomization context as it requires a complete understanding of the biological
mechanism by which each genetic marker influences the outcome. Such a priori knowledge
may be unrealistic in practice due to the possible existence of unknown
pleiotropic effects of the genetic markers \citep{little2003mendelian,Davey-Smith:2003aa,Lawlor:2008aa}. 

There has been tremendous interest in the development of methods to detect and
account for violation of IV assumptions (a)-(c), primarily in multiple-IV
settings under standard linear outcome and exposure models. Literature
addressing violations of assumption (a) is arguably the most developed and
extends to possible nonlinear models under a generalized methods of moments
framework; notable papers of this rich literature include \citet{Staiger:1997aa,stock2000gmm,Stock:2002aa,Chao:2005aa,Newey:2009aa}. A growing literature has likewise
emerged on methods to address violations of assumptions (b) and (c), a
representative sample of which includes \citet{kolesar2015identification,Bowden:2015aa,Bowden:2016aa,Kang:2016aa,hartwig2017robust,qi2019mendelian, Windmeijer:2019aa,zhao2020statistical,morrison2020mendelian,liu2020mendelian,kang2020two,tchetgen2021genius}, and \cite{ye2021genius}.  

The current paper is  most closely related to the works of \citet{Kang:2016aa,Guo:2018aa} and \cite{Windmeijer:2019aa}. Specifically,  \cite{Kang:2016aa} developed a penalized regression approach that can recover valid
inferences about the causal effect of interest provided fewer than 50\% of genetic markers are invalid instruments (known as majority rule);
\citet{Windmeijer:2019aa} improved on the penalized approach, including a
proposal for standard error estimation lacking in \citet{Kang:2016aa}. In an
alternative approach, \citet{Han:2008aa} established that the median of multiple
estimators of the effect of exposure obtained using one instrument at the time
is a consistent estimator under a majority rule and an assumption that instruments cannot have
direct effects on the outcome unless the instruments are uncorrelated. \citet{Guo:2018aa} proposed two stage hard thresholding (TSHT) with voting, which is
consistent for the causal effect under linear outcome and exposure models, and
a plurality rule which can be considerably weaker than the majority rule. The
plurality rule is defined in terms of regression parameters encoding (i) the
association of each invalid instrument with the outcome and that encoding (ii) the
association of the corresponding instrument with the exposure. The condition
effectively requires that the number of valid instruments is greater than the largest
number of invalid instruments with equal ratio of regression coefficients (i) and
(ii). Furthermore, they provide a simple construction for 95\% confidence
intervals to obtain inferences about the exposure causal effect which are guaranteed
to have correct coverage under the plurality rule.

Importantly, in these
works, a candidate instrument may be invalid either because it violates the exclusion
restriction, or because it shares an unmeasured common cause with the outcome,
i.e. either (b) or (c) fails. Both the penalized approach and the median
estimator may be inconsistent if majority rule fails, while TSHT may be
inconsistent if plurality rule fails. It is important to note that because
confidence intervals for the causal effect of the exposure obtained by \citet{Windmeijer:2019aa} and \citet{Guo:2018aa}
rely on a consistent model
selection procedure, such confidence intervals fail to be uniformly valid over
the entire model space  \citep{leeb2008sparse,Guo:2018aa}. 
Interestingly, \citet{Kang:2016aa,Guo:2018aa} and \citet{Windmeijer:2019aa}
effectively tackle the task of obtaining valid inferences about a confounded
causal effect in the presence of invalid instruments as a  model
selection problem. Specifically, they aim to correctly identify which candidate instruments are invalid, while simultaneously obtaining valid Mendelian randomization inferences using selected valid instruments, arguably a more challenging task than may be required to obtain causal inferences that are robust to the presence of
invalid instruments without necessarily knowing which instruments are invalid.

In this paper, we propose novel methods that by-pass the model selection step altogether
to deliver a regular and asymptotically linear  g-estimator \citep{robins1992estimating,robins1994estimation,vansteelandt2003causal} of the causal
parameter of interest,  provided that at least $\gamma$ out of $K$ candidate instruments are valid, for $1\leq\gamma\leq K$ set by the analyst, without
necessarily knowing their identities. A necessary trade-off
between more stringent IV\ relevance requirements in exchange for less
stringent IV causal requirements is revealed by decreasing the value of
$\gamma$. We characterize a
semiparametric efficiency bound which cannot be improved upon by any  regular and asymptotically linear 
estimator which shares the robustness property of our class of estimators. 
Although
Mendelian randomization  is used as motivating example throughout the paper,
the development of methodology to adequately address
the issue of invalid instruments remains a priority for several disciplines, including biostatistics, epidemiology, econometrics and sociology, in which our results equally apply. 

\vspace{-0.2cm}
\section{Preliminaries}
\subsection{Data and notation}
 \label{sec:notation}

Suppose that $(O_1,...,O_n)$ are independent, identically distributed observations of $O=(Y,A,\boldsymbol{Z})$ from a target population, where $Y$ is an outcome of interest, $A$ is an exposure  and $\boldsymbol{Z}=(Z_1,...,Z_K)$ comprises of  $K$  candidate instruments.  Following the tradition in the IV literature \citep{robins1994correcting,angrist1995identification,angrist1996identification, heckman1997instrumental}, to ease exposition we will focus on the canonical case of binary instruments $\boldsymbol{Z}$ which takes values in $\mathcal{Z}=\{0,1\}^K$, but the framework can be extended readily to candidate instruments with arbitrary finite support. In the context of Mendelian randomization studies, $Z_k$ represents binary coding of the minor allele variant at the $k$th single polymorphism location which  defines the $k$-th candidate instrument. 

For any index set $\alpha\subseteq \{1,...,K\}$, let $\boldsymbol{Z}_{\alpha}=(Z_s:s\in\alpha)$ and $\boldsymbol{Z}_{-\alpha}=(Z_s:s \not\in\alpha)$. Let ${\rm I\!H}$ denote the Hilbert space of one-dimensional functions of $\boldsymbol{Z}$ with mean zero and finite variance, equipped with the covariance inner product.  For a set $J$, let $|J|$ denote its cardinality. For a vector $\upsilon$, let $\upsilon^\T$ denote its transpose, $\upsilon^{\otimes 2}=\upsilon\upsilon^{\T}$, $\upsilon \odot w$ its Hadamard product with another vector $w$ of the same dimension, and $||\upsilon||_0$ its zero-norm, that is the number of nonzero elements in $\upsilon$. In addition let $\widehat{E}_n\{g(O)\}=n^{-1}\sum_{i=1}^n\{g(O_i)\}$ denote the empirical mean operator.

\subsection{Causal model}

To define the causal effects of interest under the potential outcomes framework \citep{Neyman:1923a,Rubin:1974},  let $Y(a,\mathbf{z})\in {\rm I\!R}$ denote the potential outcome that would be observed had the exposure and instruments been set to the level $a\in {\rm I\!R}$ and $\mathbf{z}\in \mathcal{Z}$ respectively, and let $A(\mathbf{z})\in {\rm I\!R}$ denote the potential exposure if the instruments would take value $\mathbf{z}\in \mathcal{Z}$. It is well known that, even when all of the candidate instruments in $\boldsymbol{Z}$ are valid,  average causal effects cannot be identified from the observed data law without further  restrictions \citep{pearl2009causality}. A common structural assumption in the invalid IV literature is the additive linear, constant effects model of {\cite{holland1988causal}} extended to allow multiple invalid instruments {\citep{small2007sensitivity}}.
\begin{assumption}
\label{assp:alice}
For two possible values of the exposure $a^{\prime}$, $a$ and a possible value of the instruments $\mathbf{z}\in\mathcal{Z}$, $$Y{(a^{\prime},\mathbf{z})}-Y{(a,\mathbf{0})}=\beta^{\ast} (a^{\prime}-a)+\psi(\mathbf{z}).$$
\end{assumption}
\noindent The unknown  parameter of interest is $\beta^{\ast}\in{\rm I\!R}$ which encodes the homogeneous causal effect per unit change in the exposure. Here $\psi(\cdot)$ is an unknown function satisfying $\psi(\mathbf{0})=0$ which encodes the direct effects of the instruments on the outcome; changing the instruments from the reference level $\mathbf{0}$ to $\mathbf{z}$ results in a direct effect of $\psi(\mathbf{z})$ on the outcome. Assumption \ref{assp:alice} rules out any interactions between the causal effect of the exposure and the direct effects of the instruments on the outcome, but allows for arbitrary interactions among the direct effects of the instruments. We formalize the definition of valid instruments in conjunction with assumption \ref{assp:alice} as follows. 
\begin{definition}
\label{def:valid}
Suppose assumption \ref{assp:alice} holds with $K$ candidate instruments. For any index set $\alpha\subseteq \{1,...,K\}$, we say that the candidate instruments   $\boldsymbol{Z}_{\alpha}$ are valid while the remaining instruments  $\boldsymbol{Z}_{-\alpha}$ are invalid if the causal model $\mathcal{C}(\alpha)$ defined by $E\{Y(0,0)|\boldsymbol{Z}\}=E\{Y(0,0)|\boldsymbol{Z}_{-\alpha}\}$ and $\psi(\boldsymbol{Z})=\psi(\boldsymbol{Z}_{-\alpha})$  holds almost surely.
\end{definition}
\noindent Definition \ref{def:valid} is closely related to the basic conditions which define valid instruments under the potential outcomes framework \citep{robins1994correcting, angrist1996identification}. Specifically, when there is only one candidate binary instrument $Z$, assumption (b) or no unmeasured confounding of the exposure-outcome relationship is typically stated as $Y(a,z)$ and $A(z)$ being  independent of $Z$ for any levels of the exposure $a$ and instrument $z$, which implies $E\{Y(0,0)|Z\}=E\{Y(0,0)\}$. In addition, exclusion restriction is typically stated as the condition $Y(a,1)=Y(a,0)$ for any level of the exposure $a$, which implies that the function $\psi(Z)=0$ does not depend on the value of $Z$. 

Suppose we have oracle knowledge of the valid and invalid instruments so that  $\mathcal{C}(\tilde{\alpha})$ is known to hold for some $\tilde{\alpha}\subseteq \{1,...,K\}$. Assumption \ref{assp:alice} yields the following semiparametric under-identified observed data model,
\begin{align}
\label{eq:semi}
Y=\beta^{\ast} A+\psi(\boldsymbol{Z})+E\{Y{(0,0)}|\boldsymbol{Z}\}+\varepsilon, \quad E(\varepsilon|\boldsymbol{Z})=0,
\end{align}
where $\varepsilon=Y{(0,0)}-E\{Y(0,0)|\boldsymbol{Z}\}$. Therefore the causal model $\mathcal{C}(\tilde{\alpha})$ implies the observed data model $\mathcal{M}(\tilde{\alpha})$ defined by the conditional mean independence restriction
\begin{align*}
E(Y-\beta^{\ast} A\mid\boldsymbol{Z}) =E(Y-\beta^{\ast} A\mid\boldsymbol{Z}_{-\tilde{\alpha}}).
\end{align*}
\noindent G-estimators of $\beta$ developed in the context of additive
and multiplicative structural mean models \citep{robins1992estimating,robins1994estimation} may be constructed in model $\mathcal{M}(\tilde{\alpha})$ based on the unconditional form
\begin{align}
\label{eq:g}
 E\{d(\boldsymbol{Z})(Y-\beta^{\ast} A)\}=E\{d(\boldsymbol{Z})E(Y-\beta^{\ast} A|\boldsymbol{Z})\}=E\{d(\boldsymbol{Z})E(Y-\beta^{\ast} A|\boldsymbol{Z}_{-\tilde{\alpha}})\}=0,
  \end{align}
for any $d(\boldsymbol{Z})\in \{h(\boldsymbol{Z}): E\{h(\boldsymbol{Z})|\boldsymbol{Z}_{-\tilde{\alpha}}\}=0\}\cap{{\rm I\!H}}$. In the absence of oracle knowledge about $\tilde{\alpha}$, identification and estimation of $\beta$ has been an area of active research \citep{kolesar2015identification, bowden2016consistent, Kang:2016aa,Windmeijer:2019aa,Guo:2018aa}. The next section briefly discusses these references to motivate our proposed approach.

\begin{remark}

Assumption \ref{assp:alice} may be weakened to accommodate heterogeneous causal effects at the individual level; see \citet{angrist1995two}, \citet{hernan2006instruments}, \citet{small2007sensitivity} and the discussion in \citet{Kang:2016aa}. For example, the observed data implication $\mathcal{C}({\alpha})\implies \mathcal{M}({\alpha})$ continue to hold if we replace assumption \ref{assp:alice} with the following additive structural mean model \citep{robins1994correcting},
$$E\{Y{(a^{\prime},\mathbf{z})}-Y{(a,\mathbf{0})}\mid \boldsymbol{Z}=\mathbf{z}\}=\beta(a^{\prime}-a)+\psi(\mathbf{z}),$$
where $\beta$ now encodes the homogeneous {\it average} additive causal effect within subpopulations defined by the values of the candidate instruments. Nonetheless, to ease exposition and comparison with prior works in the invalid IV literature, we will focus on the constant effects model in  assumption \ref{assp:alice}. A homogeneity condition akin to  assumption \ref{assp:alice} for the outcome model or a similar condition for the exposure model is necessary for point identification of average causal effects, see  \cite{robins1994correcting}, \cite{angrist1996identification}, \cite{hernan2006instruments} and \cite{wang2018bounded}. Crucially,  assumption \ref{assp:alice} is guaranteed to hold under the sharp null hypothesis of no additive causal effect, in which case a distribution-free test of the null becomes feasible. Partial identification of  causal effects with  invalid instruments  under a fully nonparametric model that does not impose  assumption \ref{assp:alice} is discussed in Section \ref{sec:partial}.
\end{remark}

\subsection{Prior work on identification and estimation of model parameters}
\label{sec:prior}
In addition to  assumption \ref{assp:alice}, \cite{Kang:2016aa}, \cite{Windmeijer:2019aa} and \cite{Guo:2018aa} assume the linear models 
 \begin{equation}
\begin{split}
\label{li}
E\{Y{(0,0)}|\boldsymbol{Z}\}=\sum_{k=1}^K s^{\ast}_k Z_k,\quad \psi(\boldsymbol{Z})=\sum_{k=1}^K t^{\ast}_k Z_k,
\end{split}
\end{equation}
which rules out interactions among the direct effects of the instruments. Violation of assumptions (b) or (c) for the $k$-th candidate instrument is encoded by $s^{\ast}_k\neq 0$ or $t^{\ast}_k\neq 0$, respectively. This definition of valid instruments is a special case of definition \ref{def:valid} and yields the under-identified single-equation linear model in \citet{wooldridge2010econometric},
 \begin{equation}
\begin{split}
\label{lout}
Y=\beta^{\ast} A+\sum_{k=1}^K \underbrace{(s^{\ast}_k+t^{\ast}_k)}_{=\pi^{\ast}_k} Z_k+ \varepsilon_1,\quad E(\varepsilon_1|\boldsymbol{Z})=0.
\end{split}
\end{equation}
Thus $\pi^{\ast}_k=0$ if the $k$-th candidate instrument is valid, and $\tilde{\alpha}=\{k:\pi^{\ast}_k=0\}$. In conjunction with the linear exposure model
 \begin{equation}
\begin{split}
\label{eq:exposure}
A=\sum_{k=1}^K \xi^{\ast}_k Z_k + \varepsilon_2,\quad E( \varepsilon_2|\boldsymbol{Z})=0,
\end{split}
\end{equation}
where $(\varepsilon_1,\varepsilon_2)$ may be correlated due to potential unmeasured confounding, the reduced form for (\ref{lout}) is
 \begin{equation*}
\begin{split}
Y=\sum_{k=1}^K \underbrace{(\beta^{\ast}\xi^{\ast}_k+\pi^{\ast}_k)}_{=\Gamma^{\ast}_k} Z_k+ (\varepsilon_1+\beta\varepsilon_2),\quad E(\varepsilon_1+\beta\varepsilon_2|\boldsymbol{Z})=0.
\end{split}
\end{equation*}
The parameters $\{\xi^{\ast}_k,\Gamma^{\ast}_k: k=1,...,K\}$ are identified and can be estimated by regressing $A$ and $Y$ on the candidate instruments respectively. Therefore identification of $\beta^{\ast}$ entails finding conditions on the model parameter space so that there is a unique, invertible mapping between $\{\beta^{\ast},\pi^{\ast}_k: k=1,...,K\}$ and $\{\xi^{\ast}_k,\Gamma^{\ast}_k:k=1,...,K\}$ through the relationship $\Gamma^{\ast}_k=\beta^{\ast}\xi^{\ast}_k+\pi^{\ast}_k$. Assuming all the candidate instruments are relevant in the sense that $|\{k:\xi^{\ast}_k\neq 0\}|=K$, a necessary and sufficient condition is that the valid instruments form a   plurality defined by the parameter values,  $|\{k:\pi^{\ast}_k= 0\}|>\max_{c\neq 0}|\{k:\pi^{\ast}_k/\xi^{\ast}_k=c\}|$ \citep{hartwig2017robust,Guo:2018aa}. A sufficient condition is $|\{k:\pi^{\ast}_k= 0\}|\geq \lceil K/2 \rceil$, also known as the majority rule \citep{Kang:2016aa,Windmeijer:2019aa}. Under these identifying conditions, the parameters $\{\beta^{\ast},\pi^{\ast}_k:k=1,...,K\}$ can be jointly estimated using penalized regression methods, so that $\tilde{\alpha}$ can in principle be recovered. Consistency of these selection procedures generally also depends on the parameter space to be  ``well-separated"  \citep{zhao2006model,https://doi.org/10.1111/j.1467-9868.2011.00771.x,Kang:2016aa,Windmeijer:2019aa,Guo:2018aa}.

\section{Multiply robust identification and estimation with invalid instruments}
\label{sec:indentification}

\subsection{Framework}

The majority rule is a popular condition that is conceptually easy to interpret and widely adopted \citep{bowden2016consistent, Kang:2016aa,Windmeijer:2019aa}.  A natural extension of interest is to identify $\beta^{\ast}$ when at least $\gamma$ out of the $K$ candidate instruments are valid, for any integer value $1\leq\gamma\leq K$ specified by the analyst. The value of $\gamma$ may
be determined by subject matter expertise. For example, a geneticist can provide a rough gauge on
the maximum number of invalid instruments in Mendelian randomization studies. Alternatively, even when such {\it a
priori} knowledge is lacking, one may wish to conduct 
sensitivity analysis in which a consistent estimator is obtained assuming at least $\gamma$ valid
instruments, over a range of values for $\gamma$ to reflect uncertainty about its true value \citep{rosenbaum2010design}.

A key challenge for pursuing in this direction under the linear causal and exposure  models discussed in section \ref{sec:prior} is that identification of $\beta^{\ast}$ is entwined with parameter values which also encode the validity of instruments. For inference, the recent work by \cite{kang2020two} establishes valid 
confidence intervals for $\beta^{\ast}$ when at least $\gamma$ instruments are valid under linear causal model (\ref{li}), by taking unions of the
confidence intervals constructed over all possible valid instrument sets $\alpha\in\{\ell \subseteq\{1,...,K\}:|\ell|= \gamma\}$. However, identification of $\beta$ may fail even when the analyst has a significant degree of {\it ex ante} confidence that at least $\gamma$ out of the $K$ candidate instruments are valid, for some $\gamma<K/2$, if the plurality rule is violated. This represents a significant gap in the invalid IV literature, which has essentially restricted identification and estimation of causal effects to $\gamma=\lceil K/2 \rceil$. This paper is the first to establish identification of $\beta^{\ast}$ in a union of multiple causal models under assumption \ref{assp:alice}, which formalizes one's prior belief that at least $\gamma$ out of the $K$ candidate instruments are valid, without {\it ex ante} knowledge about which are. Specifically, given a collection $\{\mathcal{A}({\alpha}):\alpha\in\mathcal{Q}\}$ of causal or observed data models indexed by elements of a finite set $\mathcal{Q}$, let $\cup_{\alpha\in\mathcal{Q}}\mathcal{A}(\alpha)$ denote the union model in which at least one of the models in  $\{\mathcal{A}({\alpha}):\alpha\in\mathcal{Q}\}$ holds. 

\begin{definition}
\label{def:union1}
Suppose assumption \ref{assp:alice} holds with $K$ candidate instruments. We say that { at least } $\gamma$ out of the $K$ candidate instruments are valid if the union causal model 
$\cup_{\alpha\in\{\ell \subseteq\{1,...,K\}:|\ell|\geq \gamma\}}\mathcal{C}(\alpha)$ holds.
\end{definition}

\noindent  Identification of $\beta^{\ast}$ is accomplished by exploiting the implication on the observed data, which is given in the next result. 

\begin{proposition}
\label{prop:implication}
Suppose assumption \ref{assp:alice} holds. The union causal model $\cup_{\alpha\in\{\ell \subseteq\{1,...,K\}:|\ell|\geq \gamma\}}\mathcal{C}(\alpha)$ implies the union observed data model $\cup_{\alpha\in\{\ell \subseteq\{1,...,K\}:|\ell|= \gamma\}}\mathcal{M}(\alpha)$.
\end{proposition}

 \subsection{Identification and semiparametric efficiency bound}
The causal parameter of interest $\beta^{\ast}$ is identified in the union observed data model $\cup_{\alpha\in\{\ell \subseteq\{1,...,K\}:|\ell|= \gamma\}}\mathcal{M}(\alpha)$ if it is identified in $\mathcal{M}({\alpha})$ {for each} $\alpha\in\{\ell \subseteq\{1,...,K\}:|\ell|= \gamma\}$. The key idea of our proposed approach is that the unconditional moment condition (\ref{eq:g}) holds in $\mathcal{M}({\alpha})$ {for each} $\alpha\in\{\ell \subseteq\{1,...,K\}:|\ell|= \gamma\}$, for any index function that lies in the subspace $$d(\boldsymbol{Z})\in{\rm I\!H}_{\gamma}=\{h(\boldsymbol{Z}): E\{h(\boldsymbol{Z})|\boldsymbol{Z}_{-\alpha}\}=0\text{ for each }\alpha\in\{\ell \subseteq\{1,...,K\}:|\ell|= \gamma\}\}\cap{{\rm I\!H}}.$$ 
To ease exposition in characterizing the subspace ${\rm I\!H}_{\gamma}$, we first assume joint independence of the candidate instruments which we later relax in Section \ref{sec:corr}.
\begin{assumption}
 \label{assp:indp}
The distribution of the  candidate instruments  is non-degenerate and factorizes as  $\Pr(\boldsymbol{Z}\leq\mathbf{z})=\prod_{s=1}^K \Pr(Z_s\leq z_s)$ for any $\mathbf{z}\in\mathcal{Z}$.
\end{assumption}
\noindent Assumption \ref{assp:indp} may be reasonable in Mendelian randomization studies by performing  linkage disequilibrium clumping to choose an independent set of genetic variants as candidate instruments using the well-established genetics analysis software PLINK \citep{purcell2007plink}. 

\begin{example}
\label{ex:1}
To fix ideas, suppose assumption \ref{assp:indp} holds with $K=2$ candidate instruments. Then ${\rm I\!H}$ can be spanned by a set of $2^2-1$ orthogonal functions,
\begin{align*}
{\rm I\!H}&=\text{span}(\{Z_1-\mu^{\ast}_1,Z_2-\mu^{\ast}_2,(Z_1-\mu^{\ast}_1)(Z_2-\mu^{\ast}_2)\}),
\end{align*}
where $\mu^{\ast}_k=E(Z_k)$ for $k=1,...,K$. In addition, suppose the analyst has a significant degree of {\it ex ante} confidence that at least $\gamma=1$ of the candidate instruments is valid, in which case we hope to characterize the subspace ${\rm I\!H}_{1}=\{h(\boldsymbol{Z}): E\{h(\boldsymbol{Z})|\boldsymbol{Z}_{-\alpha}\}=0\text{ for each }\alpha\in\{\{1\},\{2\}\}\}\cap{{\rm I\!H}}=\{h(\boldsymbol{Z}): E\{h(\boldsymbol{Z})|\boldsymbol{Z}_{k}\}=0\text{ for each }k=1,2\}\cap{{\rm I\!H}}$. It is straightforward to verify that only the demeaned two-way interaction $(Z_1-\mu^{\ast}_1)(Z_2-\mu^{\ast}_2)$ satisfies $E\{(Z_1-\mu^{\ast}_1)(Z_2-\mu^{\ast}_2)\mid Z_k\}=0$ for each $k=1,2$. Therefore, the moment condition
$$E\{(Z_1-\mu^{\ast}_1)(Z_2-\mu^{\ast}_2)(Y-\beta)\}=0,$$
holds in $\mathcal{M}(\alpha)$ for each $\alpha=\{\{1\},\{2\}\}$, that is in the union observed data model $\cup_{\alpha\in\{\{1\},\{2\}\}}\mathcal{M}(\alpha)$.  Since ${\rm I\!H}=\text{span}(\{Z_1-\mu^{\ast}_1,Z_2-\mu^{\ast}_2\})\oplus\text{span}(\{(Z_1-\mu^{\ast}_1)(Z_2-\mu^{\ast}_2)\})$, we can conclude that ${\rm I\!H}_{1}=\text{span}(\{(Z_1-\mu^{\ast}_1)(Z_2-\mu^{\ast}_2)\})=\{\theta (Z_1-\mu^{\ast}_1)(Z_2-\mu^{\ast}_2):\theta\in {\rm I\!R}\}$.
\end{example}
Intuitively, because at most $K-\gamma$ of the candidate instruments are invalid, each of the demeaned interactions of order at least $K-\gamma+1$ necessarily involves at least one valid instrument, without {\it ex ante} knowledge about their identities. We generalize this result for arbitrary values of $K$ and $1\leq\gamma\leq K$ below .

{
 \begin{theorem}
 \label{prop:mr1}
Suppose assumption \ref{assp:indp} holds with $K$ candidate instruments. Then 
$${\rm I\!H}_{\gamma}=\text{span}\left(\left\{\Pi_{s\in\alpha}(Z_s-\mu^{\ast}_s):\alpha\in\{\ell \subseteq\{1,...,K\}:K-\gamma+1\leq |\ell|\leq K\}\right\}\right),$$
and accordingly the dimension of ${\rm I\!H}_{\gamma}$ is $d_{\gamma}=\sum_{j=K-\gamma+1}^K{K \choose j}$. Without loss of generality, we consider the enumeration $\{\alpha(1),...,\alpha(d_{\gamma})\}$ of the elements in $\left\{\ell \subseteq\{1,...,K\}:K-\gamma+1\leq |\ell|\leq K\}\right\}$ in some fixed order and let $\mathbf{D}_{\gamma}(\boldsymbol{Z};\boldsymbol{\mu}^{\ast})=\{\Pi_{s\in\alpha(1)}(Z_s-\mu^{\ast}_s),...,\Pi_{s\in\alpha(d_{\gamma})}(Z_s-\mu^{\ast}_s)\}^\T$ where $\boldsymbol{\mu}^{\ast}=(\mu^{\ast}_1,...,\mu^{\ast}_K)^\T$.  Then the causal parameter of interest $\beta^{\ast}$ is identified in the union observed data model $\cup_{\alpha\in\{\ell \subseteq\{1,...,K\}:|\ell|= \gamma\}}\mathcal{M}(\alpha)$ as the unique solution to the $d_{\gamma}$ unconditional moment restrictions
\begin{align}
\label{eq:ident}
E\{\mathbf{D}_{\gamma}(\boldsymbol{Z};\boldsymbol{\mu}^{\ast})(Y-\beta A)\}=\mathbf{0},
\end{align}
provided that 
\begin{align}
\label{eq:ident2}
\biggr|\biggr| \frac{\partial E\{\mathbf{D}_{\gamma}(\boldsymbol{Z};\boldsymbol{\mu}^{\ast})(Y-\beta A)\}}{\partial \beta}\biggr|\biggr| _0=||E\{\mathbf{D}_{\gamma}(\boldsymbol{Z};\boldsymbol{\mu}^{\ast})A\}||_0> 0.
\end{align}
\end{theorem}
Condition (\ref{eq:ident2}) requires at least one of the $d_{\gamma}$ functions in $\mathbf{D}_{\gamma}(\boldsymbol{Z};\boldsymbol{\mu}^{\ast})$ to be correlated with the exposure, which is a rank condition for identification with invalid instruments. Note that this requirement becomes less stringent as $\gamma$ increases. In fact, if we believe that all of the $K$ candidate instruments are valid, then we need at least one out of the $d_{K}=2^K-1$ main and interaction terms to be correlated with the exposure. The semiparametric identification framework outlined in Theorem  \ref{prop:mr1} is particularly instructive as it reveals a fundamental trade-off between the strength of required rank condition, and the potential number of valid instruments one is willing to allow in a given analysis. In practice, appropriateness of the IV relevance assumption in question for a given value of $\gamma$  is empirically testable, as later illustrated in the UK Biobank data analysis in Section \ref{sec:UKB}.

\begin{remark}

In conjunction with Proposition \ref{prop:implication}, Theorem \ref{prop:mr1} establishes multiply robust causal identification,  because $\beta^{\ast}$ is identified as long as one of the causal models in the collection $\{\mathcal{C}({\alpha}):\alpha\in\{\ell \subseteq\{1,...,K\}:|\ell|\geq \gamma\}\}$ holds. The multiple robustness result of Theorem \ref{prop:mr1}  is distinct from prior literature which primarily concerns nuisance model specification under a single causal model \citep{10.2307/2669930,vansteelandt2008multiply,tchetgen2012semiparametric,molina2017multiple,babino2019multiple}, whereas our  notion of multiply robust inference concerns identification in the union of causal models. To the best of our knowledge, Theorem \ref{prop:mr1} constitutes the first instance of multiple robustness property for causal identification.
\end{remark}

In this paper, we consider regular and asymptotically linear estimators of $\beta^{\ast}$ implicitly defined by the $d_{\gamma}$ moment conditions (\ref{eq:ident}), in the semiparametric model only restricted by assumption \ref{assp:indp} and condition (\ref{eq:ident2}). An estimator is regular in the semiparametric model if it converges to its limiting distribution in a locally uniform manner in any regular parametric submodel therein; see \cite{bickel1993efficient} and \cite{newey1990semiparametric} for more precise definitions. Regularity of an estimator is often considered desirable, as the estimator's limiting distribution is not affected by small vanishing perturbations of the data generating process and thus the resulting confidence intervals are uniformly valid over the entire model space. 

\begin{theorem}
 \label{prop:mr2}
 Let $\beta^{\ast}$ be defined by the $d_{\gamma}$ moment conditions (\ref{eq:ident}). Under assumption \ref{assp:indp} and condition (\ref{eq:ident2}), any regular and asymptotically linear estimator $\widehat{\beta}(\gamma)$ of $\beta^{\ast}$ must satisfy the following asymptotic expansion:
\begin{align*}
 {n}^{1/2} \{ \hat{\beta}(\gamma)-\beta^{\ast}\}  = E\{\Theta^\T\mathbf{D}_{\gamma}(\boldsymbol{Z};\boldsymbol{\mu}^{\ast})A\}^{-1} {n}^{1/2}\widehat{E}_n%
\{\Theta^\T\mathbf{G}_{\gamma}(O;\beta^{\ast},\boldsymbol{\mu}^{\ast})\}+o_{p}\left(  1\right)  ,   
\end{align*}
for any $d_{\gamma}$-dimensional real vector $\Theta$ that satisfies $E\{\Theta^\T\mathbf{D}_{\gamma}(\boldsymbol{Z};\boldsymbol{\mu}^{\ast})A\}\neq 0$, where $\mathbf{G}_{\gamma}(O;\beta,\boldsymbol{\mu})=\mathbf{D}_{\gamma}(\boldsymbol{Z};\boldsymbol{\mu})\odot \{(Y-\beta A)+\mathbf{A}_{\gamma}(\boldsymbol{Z};\beta)\}$ and $\mathbf{A}_{\gamma}(\boldsymbol{Z};\beta)$ is an  $d_{\gamma}$-dimensional augmentation vector with the $j$-th entry equal to
\begin{align*}
\sum_{1\leq r \leq |\alpha(j)|}\left\{\sum_{\alpha\in\{\ell\subseteq\alpha(j):|\ell|=r\}}(-1)^{r} E(Y-\beta A\mid \boldsymbol{Z}_{-\alpha})\right\}.
\end{align*}
Furthermore, the efficient regular and asymptotically linear estimator estimator admits the above expansion with $\Theta_{opt}=E\{\mathbf{G}^{\otimes 2}_{\gamma}(O;\beta^{\ast},\boldsymbol{\mu}^{\ast})\}^{-1}E\{\mathbf{D}_{\gamma}(\boldsymbol{Z};\boldsymbol{\mu}^{\ast})A\}$, so that the asymptotic variance lower bound for all regular estimators of $\beta^{\ast}$ is given by
$$\mathcal{V}_{\gamma}=[E\{A\mathbf{D}^\T_{\gamma}(\boldsymbol{Z};\boldsymbol{\mu}^{\ast})\} E\{\mathbf{G}^{\otimes 2}_{\gamma}(O;\beta^{\ast},\boldsymbol{\mu}^{\ast})\}^{-1}E\{\mathbf{D}_{\gamma}(\boldsymbol{Z};\boldsymbol{\mu}^{\ast})A\}]^{-1}.$$

\end{theorem}

{Because $\mathbf{D}_{\gamma}(\boldsymbol{Z};\boldsymbol{\mu}^{\ast})$ is a subvector of $\mathbf{D}_{\gamma^{\prime}}(\boldsymbol{Z};\boldsymbol{\mu}^{\ast})$ for any $\gamma < \gamma^{\prime} $, the corresponding efficiency bounds satisfy $\mathcal{V}_{\gamma}\geq \mathcal{V}_{\gamma^{\prime}}$. Theorem \ref{prop:mr2} therefore provides a formal quantification of the efficiency cost one must incur in exchange for increased causal robustness, as $\cup_{\alpha\in\{\ell \subseteq\{1,...,K\}:|\ell|\geq \gamma\}}\mathcal{C}(\alpha)$ subsumes $\cup_{\alpha\in\{\ell \subseteq\{1,...,K\}:|\ell|\geq \gamma^{\prime}\}}\mathcal{C}(\alpha)$. It also  establishes that the  IV relevance condition (\ref{eq:ident2}) is in fact necessary  for  the existence of a  regular and asymptotically linear  estimator of $\beta^{\ast}$.}

Unless the candidate instruments are under the control of the investigator through some known allocation scheme such as randomisation, their marginal means $\boldsymbol{\mu}^{\ast}$ are typically unknown in observational studies. Under assumption \ref{assp:indp}, we can obtain the nonparametric estimator $\widehat{\boldsymbol{\mu}}$ which solves
\begin{align}
\label{eq:nuis}
\widehat{E}_n\{(Z_1-\mu_1,...,Z_K-\mu_K)^\T\}=0.
\end{align}
The augmentation vector $\mathbf{A}_{\gamma}(\boldsymbol{Z};\beta)$ may be viewed as an adjustment to the influence function arising from nonparametric estimation of the nuisance parameters $\boldsymbol{\mu}^{\ast}$. To illustrate, suppose assumption \ref{assp:indp} holds with $K=2$ candidate instruments and the analyst believes that at least $\gamma=1$ of the candidate instruments is valid, in which case $\mathbf{D}_{1}(\boldsymbol{Z};\boldsymbol{\mu}^{\ast})$ comprises only of the demeaned two-way interaction $(Z_1-\mu^{\ast}_1)(Z_2-\mu^{\ast}_2)$; see Example \ref{ex:1}. By Theorem \ref{prop:mr2}, any influence function for estimation of $\beta^{\ast}$ defined by the moment condition
$E\{(Z_1-\mu^{\ast}_1)(Z_2-\mu^{\ast}_2)(Y-\beta A)\}=0$ is, up to multiplicative constants, of the form
\begin{align}
\label{eq:if}
(Z_1-\mu^{\ast}_1)(Z_2-\mu^{\ast}_2)\{Y-\beta^{\ast}A-E(Y-\beta^{\ast}A|Z_1)-E(Y-\beta^{\ast}A|Z_2)+E(Y-\beta^{\ast}A)\}.
\end{align}
Here the three outcome regressions in the augmentation term correspond to adjustments arising from nonparametric estimation of $\mu^{\ast}_1$, $\mu^{\ast}_2$ and $\mu^{\ast}_1 \mu^{\ast}_2$. Furthermore, it can be verified that the moment condition 
\begin{align}
\label{eq:alt}
E\{Y-\beta^{\ast}A-E(Y-\beta^{\ast}A|Z_1)-E(Y-\beta^{\ast}A|Z_2)+E(Y-\beta^{\ast}A)\}=0,
\end{align}
holds in $\mathcal{M}(\alpha)$ for each $\alpha\in\{\{1\},\{2\}\}$, which represents an alternative identification approach involving outcome regressions. When nonparametric methods are used to estimate these outcome regressions, estimators based on moment condition (\ref{eq:alt}) also have influence functions of the form in (\ref{eq:if}).

\subsection{G-estimation via two-stage least squares}
\label{sec:g-estimation}
\cite{newey1999two} and \cite{ackerberg2014asymptotic} showed that a form of semiparametric optimally
weighted generalized method of moments (GMM) estimator  of \cite{Hansen:1982aa} with $\widehat{\boldsymbol{\mu}}$ plugged-in achieves the efficiency bound $\mathcal{V}_{\gamma}$. In this paper, following \citet{robins2000robust} and \citet{okui2012doubly}, we  propose estimation of $\beta^{\ast}$ via the following standard two-stage least squares procedure which may be readily implemented using  existing off-the-shelf software.\\

{\it Step 1.} Regress $A$ on $\{1,\mathbf{D}^T_{\gamma}(\boldsymbol{Z};\widehat{\boldsymbol{\mu}})\}^\T$ by ordinary least squares to obtain the  fitted values
$
\widehat{A}=\widehat{\theta}_{0}+\widehat{\theta}^\T \mathbf{D}_{\gamma}(\boldsymbol{Z};\widehat{\boldsymbol{\mu}}).%
$\\

{\it Step 2.} Regress $Y$ on $(1,\widehat{A})^\T$ by ordinary least squares to obtain the estimated slope parameter $\widehat{\beta}_{g}(\gamma)$.\\

It follows from the results of \citet{robins2000robust} and \citet{okui2012doubly} that $\widehat{\beta}_g(\gamma)$ admits the asymptotic linear expansion in Theorem  \ref{prop:mr2} indexed by $\Theta_{g}=E\{\mathbf{D}^{\otimes 2}_{\gamma}(\boldsymbol{Z};\boldsymbol{\mu}^{\ast})\}^{-1}E\{\mathbf{D}_{\gamma}(\boldsymbol{Z};\boldsymbol{\mu}^{\ast})A\}$, the population least squares coefficient vector from regressing $A$ on $\mathbf{D}_{\gamma}(\boldsymbol{Z};\boldsymbol{\mu}^{\ast})$. For inference, the standard sandwich estimator of the asymptotic variance treating $\widehat{\boldsymbol{\mu}}$ as known is conservative \citep{robins1992estimating,vansteelandt2003causal}, albeit readily available as an output from existing  software which implements two-stage least squares. A convenient alternative for obtaining standard errors is to implement the nonparametric bootstrap which appropriately accounts for all sources of uncertainty. 
}

\section{Simulation Studies}
\label{sec:simulation}
\subsection{Simulation design}

We perform Monte Carlo simulations to investigate the numerical performance of the proposed g-estimation method. The instruments $\boldsymbol{Z}=(Z_1,...,Z_K)^{\T}$ are generated from independent Bernoulli distributions with probability $p=0.8$. Similar to the study designs in \citet{Guo:2018aa} and \citet{Windmeijer:2019aa}, the outcome and exposure are generated without covariates from 
\begin{align*}
Y&=\beta A +  \pi^{\T} \boldsymbol{Z}+\varepsilon_1,\\
A&= \xi^{\T} f(\boldsymbol{Z})+\varepsilon_2,
\end{align*} 
where $f(\boldsymbol{Z})$ is a vector comprising of all main and interaction terms of $\boldsymbol{Z}$, and
\begin{align*}
\begin{pmatrix}\varepsilon_1\\
\varepsilon_2
\end{pmatrix} \sim  N
\begin{bmatrix}
\begin{pmatrix}
0\\
0
\end{pmatrix}\!\!,&
\begin{pmatrix}
1 & 0.25 \\
0.25 & 1 
\end{pmatrix}
\end{bmatrix}.
\end{align*} 
We set the number of candidate instruments to be $K=5$, the causal effect parameter $\beta=1$ and vary (1) the sample size $n=10000$ or $50000$; (2) the number of valid instruments given by $K-||\pi||_0$; and (3) the IV strength $\xi=(1,...,1)^TC$ for $C=0.6$ or $1$. Under the above data generating mechanism, the  coefficients indexing the linear model $A={\xi}^{\T} \boldsymbol{Z}+{\varepsilon_2}$ which omits interactions also equals ${\xi}=(1,...,1)^T{C}^{\prime}$ for some constant ${C}^{\prime}$. For (2), we first set $\pi=(0,0,0,0.2,0.2)^T$ encoding three valid instruments so that the majority rule of \citet{Kang:2016aa} holds. Then we consider the case where only two instruments are valid, and set $\pi=(0,0,0.1,0.2,0.3)^T$ or $\pi=(0,0,0.2,0.2,0.2)^T$. The plurality rule of \citet{Guo:2018aa} holds in the former design but is violated in the latter. 

We use the R package \texttt{AER} \citep{AppliedEconometricswithR} to implement  $\widehat{\beta}_{g}(2)$ (g-esimator).  A sandwich estimate of the asymptotic variance is available directly as an output from the function \texttt{ivreg}. We compare g-esimator with the oracle two stage least squares (2SLS) estimator using only the valid instruments,  the naive 2SLS estimator using $\boldsymbol{Z}$, the post-adaptive Lasso (Post-Lasso) estimator of \citet{Windmeijer:2019aa} which uses adaptive Lasso tuned with an initial median estimator to select valid instruments, and the two-stage hard thresholding (TSHT) estimator of \citet{Guo:2018aa}. The oracle 2SLS estimator is included as a benchmark. It requires {\it a priori} knowledge of which of the putative instruments are valid, and thus is infeasible in practice. 

\subsection{Monte Carlo results}

Table \ref{tab:ident} summarizes the estimation results for the setting with $C=0.6$ based on 1000 replications. When the majority rule holds, both Post-Lasso and TSHT perform nearly as well as the oracle 2SLS in terms of absolute bias, coverage and variance.  When the majority rule is violated but the plurality rule holds, TSHT performs as well as the oracle 2SLS once $n=50000$. As Post-Lasso relies on the majority rule condition for consistency, it performs poorly in this setting. When both the majority and the plurality rules are violated, Post-Lasso and TSHT exhibit noticeable bias and poor coverage. In agreement with theory, naive 2SLS suffers from poor coverage throughout, regardless of sample size. The proposed  g-estimator performs well across all simulation settings, including those in which both the majority and the plurality rules are violated, but with higher variance than the other methods. The absolute bias and variance of the estimators are generally smaller with stronger IV when $C=1$, see the Supplementary Material for estimation results of this setting.

\subsection{Further Monte Carlo results}

The saturated linear exposure model in the simulation study was chosen to ease comparison with existing methods. The Supplementary Material also contains Monte Carlo results under two additional data generating mechanisms of interest, namely (i) sparse linear exposure models in which not all the interaction terms are present, and (ii) semiparametric single index models $E(A|\boldsymbol{Z})=g(\xi^{\T}\boldsymbol{Z})$ where $g(\cdot)$ is a link function. The proposed g-estimator is able to deliver the desired level of  confidence interval coverage across the range of simulation settings considered, in line with our theory. Notably, the other competing methods all fail to have correct confidence interval coverage when both the majority and plurality rules are violated.

  \begin{table}

		\begin{center}
		\caption{Comparison of methods with a continuous exposure. The
two rows of results for each estimator correspond to sample sizes of $n = 10000$ and $n = 50000$
respectively.}
	\label{tab:ident}
		\bigskip
	
		\begin{tabular}{cccccccccccc}
			\toprule
	    & G-estimator & Oracle 2SLS & Naive 2SLS &Post-lasso & TSHT \\
	     \hline\noalign{\smallskip}
&  \multicolumn{5}{c}{Majority rule holds}\\
$|\text{Bias}|$ & 0.007 & 0.000& 0.014 & 0.000 & 0.000 \\
                      & 0.002 & 0.000 & 0.014 & 0.000 & 0.000 \\
$\sqrt{\text{Var}}$ & 0.019 & 0.002 & 0.002 & 0.003 & 0.003  \\   
                             &0.009 & 0.001 & 0.001 & 0.001 & 0.002   \\                
$\sqrt{\text{EVar}}$ & 0.019 & 0.002 & 0.002 & 0.003 & 0.003 \\
                               & 0.009 & 0.001 & 0.001 & 0.001 & 0.001 \\
Cov95 & 0.929 & 0.951 & 0.000 & 0.940 & 0.950 \\
          & 0.943 & 0.945 & 0.000 & 0.947 & 0.947 \\

&  \multicolumn{5}{c}{Majority rule violated but plurality rule holds}\\

$|\text{Bias}|$ & 0.032 & 0.000& 0.021 & 0.014 & 0.009 \\
                      & 0.017 & 0.000 & 0.021 & 0.012 & 0.000 \\
$\sqrt{\text{Var}}$ & 0.058 & 0.003 & 0.002 & 0.009 & 0.009   \\   
                             &0.043 & 0.001 & 0.001 & 0.007 & 0.003    \\                
$\sqrt{\text{EVar}}$ & 0.068 & 0.003 & 0.002 & 0.003 & 0.003  \\
                               & 0.047 & 0.001 & 0.001 & 0.001 & 0.001  \\
Cov95 & 0.935 & 0.938 & 0.000 & 0.118 & 0.424 \\
          & 0.933 & 0.951 & 0.000 & 0.000 & 0.944 \\ 
&  \multicolumn{5}{c}{Both majority and plurality rules  violated}\\
$|\text{Bias}|$ & 0.032 & 0.000& 0.021 & 0.035 & 0.035 \\
                      & 0.017 & 0.000 & 0.021 & 0.036 & 0.035 \\
$\sqrt{\text{Var}}$ & 0.057 & 0.003 & 0.002 & 0.003 & 0.003  \\   
                             &0.043 & 0.001 & 0.001 & 0.001 & 0.002   \\                
$\sqrt{\text{EVar}}$ & 0.068 & 0.003 & 0.002 & 0.002 & 0.002 \\
                               & 0.047 & 0.001 & 0.001 & 0.001 & 0.001 \\
Cov95 & 0.933 & 0.941 & 0.000 & 0.000 & 0.000 \\
          & 0.933 & 0.950 & 0.000 & 0.000 & 0.000 \\
\hline
		\end{tabular}
   \begin{tablenotes}
      \item {\noindent \small Note: $|\text{Bias}|$ and $\sqrt{\text{Var}}$ are the Monte Carlo absolute bias and standard deviation of the point estimates,  $\sqrt{\text{EVar}}$ is the
square root of the mean of the variance estimates and Cov95 is the coverage proportion of
the 95\% confidence intervals, based on 1000 repeated simulations. Zeros denote values smaller than $0.0005$.}
    \end{tablenotes}
	\end{center}
  \end{table}

\section{An Application to the UK Biobank Data}
\label{sec:UKB}

 The UK Biobank  is a large-scale ongoing prospective cohort study that recruited around 500,000 participants aged 40-69 at recruitment from 2006 to 2010. Participants provided biological samples, completed questionnaires, underwent assessments, and had nurse led interviews. Follow up is chiefly through cohort-wide linkages to National Health Service data, including electronic, coded death certificate, hospital, and primary care data \citep{sudlow2015uk}.  To reduce confounding bias due to population stratification, we restrict our analysis to people of genetically verified white British descent, as in previous studies \citep{Tyrrell2016}. We are interested in estimating the causal effects of body mass index (BMI) on systolic blood pressure (SBP) and diastolic blood pressure (DBP) respectively. Participants who are taking  anti-hypertensive medication based on self-report are excluded; the sample size for the final analysis is 292,757. 
We use top 10 independent single nucleotide polymorphisms (SNPs) that are strongly associated with BMI at genome-wide significance level $p < 5\times 10^{-8}$ \citep{Locke:2015aa}. We then implement the proposed g-estimator $\widehat{\beta}_{g}(\gamma)$ over a range of values for $\gamma$, along with the following methods: Naive two stage least squares (2SLS), TSHT \citep{Guo:2018aa}, sisVIVE \citep{Kang:2016aa} and Post-lasso \citep{Windmeijer:2019aa}. 

The data analysis results are summarized in Table \ref{tab:ukb}. For the BMI-SBP relationship, TSHT selected all 10 SNPs as valid instruments, which explains why its point estimate is similar to naive 2SLS and $\widehat{\beta}_{g}(10)$. Interestingly for the BMI-DBP relationship, $\widehat{\beta}_{g}(4)$ delivers a point estimate of $0.649$, which is larger than those of competing methods. When $\gamma=6$ or 8 and thus more than half of the SNPs are assumed to be valid, the point estimate of $\widehat{\beta}_{g}(\gamma)$ is similar to those of  sisVIVE  and Post-lasso, both of which require the 50\% rule for consistency.  We also observe that the standard errors of $\widehat{\beta}_{g}(\gamma)$ decrease substantially when the value of  $\gamma$ increases, which is in agreement with  theory.  In particular, when $\gamma \leq 3$, the $p$-values for the F-test of the first stage regression for weak instruments is not significant at $0.05$ level, and hence the point estimates  are likely to be biased due to  weak instruments. For this reason, we do not report results for $\gamma \leq 3$.  

We further perform the Hausman homogeneity test \citep{hausman1978specification} to compare the point estimates of $\widehat{\beta}_{g}(\gamma)$ when $\gamma = 6$, $8$ or $10$ versus that of $\widehat{\beta}_{g}(4)$. Under the null hypothesis $H_0: \gamma=4$, $\widehat{\beta}_{g}(4)$ is efficient under homogeneity restrictions while $\widehat{\beta}_{g}(\gamma)$ remains consistent for $\gamma> 4$. For the BMI-SBP relationship, we did not find statistically significant differences at $0.05$ level for all three pairwise tests. For the BMI-DBP relationship, we find that the point estimates of $\widehat{\beta}_{g}(\gamma)$ for $\gamma=8$ or 10 are statistically different  from that of $\widehat{\beta}_{g}(4)$ at $0.05$ level, which suggests that inferences under $\gamma=4$ may be most reliable. Nonetheless, we suggest reporting results as in Table \ref{tab:ukb} over a range of values for $\gamma$ to reflect uncertainty about its true value \citep{rosenbaum2010design}.

\begin{table}[!htbp]
	\caption{Point estimates and standard errors (in parenthesis) of the causal effects of body mass index on systolic blood pressure and diastolic blood pressure in the UK Biobank of European descent ($K=10, n=292,757)$.}
		\label{tab:ukb}
		\begin{center}
		\begin{tabular}{lcccccccc} 
		\hline 
        \rule{0pt}{3ex}  
        & Naive 2SLS & TSHT & sisVIVE  & Post-lasso   &  $\widehat{\beta}_{g}(4)$ &   $\widehat{\beta}_{g}(6)$ & $\widehat{\beta}_{g}(8)$ & $\widehat{\beta}_{g}(10)$\\ \hline

 &  \multicolumn{8}{c}{Systolic blood pressure}\\
 & 0.447     & 0.447& 0.604 & 0.628   & 0.543 & 0.523   & 0.530 & 0.496   \\
  & (0.109)      & (0.109) & $-$ &(0.133) & (0.296) & (0.135)   & (0.090) & (0.090)  \\
 \rule{0pt}{3ex}  &  \multicolumn{8}{c}{Diastolic blood pressure}\\
 &  0.254 & 0.296 & 0.386 & 0.459  & 0.649 & 0.410 & 0.363 & 0.349 \\
 & (0.059) & (0.059) & $-$ & (0.065)  &(0.147) & (0.071) & (0.048)  &(0.048)\\ 
 \hline 
 \multicolumn{9}{l}{}
\end{tabular}
\end{center}
\end{table}

{
\section{Extensions}
\label{sec:extensions}

\subsection{Correlated and continuous candidate instruments}
\label{sec:corr}
 The results from Section \ref{sec:indentification} can be generalized for arbitrary distribution of the candidate instruments.
\begin{proposition}
\label{prop3}
Suppose there are $K$ candidate instruments with joint probability mass function $\Pr(\mathbf{Z}=\mathbf{z})$ and marginal probability mass function $\Pr(Z_k=z_k)$ for the $k$-th instrument, $k=1,...,K$. Then the parameter $\beta^{\ast}$ is identified in the union observed data model $\cup_{\alpha\in\{\ell \subseteq\{1,...,K\}:|\ell|= \gamma\}}\mathcal{M}(\alpha)$ as the unique solution to the unconditional moment restrictions
\begin{align}
\label{eq:ident3}
E\{W(\boldsymbol{Z})\mathbf{D}_{\gamma}(\boldsymbol{Z};\boldsymbol{\mu}^{\ast})(Y-\beta A)\}=\mathbf{0},
\end{align}
where $W(\mathbf{z})=\prod_{s=1}^K Pr(Z_k=z_k)/\Pr(\mathbf{Z}=\mathbf{z})$, provided that 
\begin{align}
\label{eq:ident4}
\biggr|\biggr| \frac{\partial E\{W(\boldsymbol{Z})\mathbf{D}_{\gamma}(\boldsymbol{Z};\boldsymbol{\mu}^{\ast})(Y-\beta A)\}}{\partial \beta}\biggr|\biggr| _0=||E\{W(\boldsymbol{Z})\mathbf{D}_{\gamma}(\boldsymbol{Z};\boldsymbol{\mu}^{\ast})A\}||_0> 0.
\end{align}
\end{proposition}
\noindent The weights $W(\boldsymbol{Z})$ formally accounts for possible dependence among the candidate instruments and in fact reduces to $1$ for all units under assumption \ref{assp:indp}, recovering the result in Theorem \ref{prop:mr1}. When $\boldsymbol{Z}$ contains continuous components,  ${\rm I\!H}_{\gamma}$ is an infinite dimensional Hilbert space. We can adopt the general strategy proposed in \citet{NEWEY1993419} (see also \citet{tchetgen2010doubly}) by taking a basis system $\{\phi_{j}(\boldsymbol{Z}),j=1,2,...\}$ of functions dense in  ${\rm I\!H}_{\gamma}$, such as tensor products of trigonometric, wavelets or polynomial bases. In practice we can perform two stage least squares using the empirical version of $[\phi_{1}(\boldsymbol{Z}),...,\phi_{J}(\boldsymbol{Z})]^\T$ as the vector of instruments, for some $J< \infty$.

\subsection{Measured baseline covariates}

In many applications the IV assumptions may be more credible only after conditioning on a sufficiently rich set of measured baseline covariates $\boldsymbol{X}$ but not otherwise, in the sense that within strata of $\boldsymbol{X}$, the candidate instruments may be viewed as being randomized through some natural or quasi-experiment \citep{hernan2006instruments}. {In the Supplementary Material, we provide the conditional analogues of Theorems \ref{prop:mr1} and \ref{prop:mr2} under such settings. When $\boldsymbol{X}$ comprises of several variables with finite support, saturated covariate main effects can be specified for the marginal distributions of the candidate instruments given covariates, as well as for the exposure and outcome regressions in the two-stage least squares procedure of Section \ref{sec:g-estimation}. This becomes infeasible in moderately sized samples when $\boldsymbol{X}$ is high dimensional or contains numerous continuous components as the data are too sparse to conduct stratified
estimation \citep{robins1997toward}. One dimension reduction strategy is to specify parametric working models for the covariate effects. Improved robustness against model misspecification is achieved through doubly robust estimators which remain root-$n$ consistent for the causal effect of interest $\beta^{\ast}$ if either a model for the covariate main effects on the outcome or a model
for the marginal distributions of the candidate instruments, given covariates, is correctly specified, but not necessarily both \citep{robins1994correcting,okui2012doubly,vansteelandt2018improving}. Another approach uses various data-adaptive statistical or machine learning methods to estimate the covariate effects. Recent works by \citet{Chernozhukov2018ddml,chernozhukov2016locally} established rate conditions for these flexible learners such that the resulting debiased machine learning estimators of $\beta^{\ast}$ can still be root-$n$ consistent even when the complexity of the nuisance parameter space  is no longer limited by
classical settings (e.g. Donsker classes),  by exploiting Neyman orthogonality of the influence functions which translates to reduced sensitivity under local variation in the nuisance parameter. The impact of
regularization bias and overfitting in estimation of the nuisance parameters is further mitigated via cross-fitting.
We outline the implementations of both the doubly robust as well as cross-fitted debiased machine learning estimators in the Supplementary Material.}

 \subsection{Partial identification}
 \label{sec:partial}
The proposed approach allows for partial identification of average causal effects with discrete or bounded continuous outcome  and possibly invalid instruments, even in the absence of  assumption \ref{assp:alice}. The key observation that allows one to obtain valid
bounds for a causal effect of interest assuming at least $\gamma$ out of $K$
candidate instruments are valid is that, as pointed out in Section \ref{sec:indentification}, such
assumption logically implies that all higher order interactions involving at
least $K-\gamma+1$ candidate instruments must in fact satisfy the exclusion
restriction, since such interactions can at most involve $K-\gamma$ invalid
instruments and at least one valid instrument. With these interactions defined, one may then proceed
with any existing IV bounds in the literature to obtain partial inference for
a variety of  causal effects \citep{swanson2018partial}.  The approach also readily extends to binary, count and censored survival outcome settings simply by re-coding the candidate instruments and using the derived interactions as valid instruments in appropriate IV analysis for the given type of outcome available in the causal inference literature \citep{robins1994correcting,abadie2003semiparametric,vansteelandt2003causal,tan2010marginal,tchetgen2015instrumental,wang2018bounded,liu2020identification}.  

}

\section{Discussion}
\label{sec: discussion}

In closing, we acknowledge certain limitations of the proposed semiparametric method. First, the approach may be vulnerable to bias which may occur if exposure is weakly dependent on the interactions. However, although not pursued here, in principle one can leverage existing literature on many weak instruments to address this challenge \citep{Newey:2009aa, ye2021genius}. Second, in observational studies the IV assumptions may only hold conditional on a high dimensional set of baseline covariates and the multiple candidate instruments may be dependent, thus requiring methods briefly introduced in Section \ref{sec:corr}. In such settings, the need to model the joint density of candidate instruments presents several computational and inferential challenges, which we plan to address in future work.

\section*{Acknowledgement}
 Baoluo Sun is supported by the National University of Singapore Start-Up Grant (R-155-000-203-133). Zhonghua Liu is supported by Hong Kong Research Grants Council Early Career Scheme (27307920). Eric Tchetgen Tchetgen's work is funded by NIH grant R01AI104459. This research has been conducted using the UK Biobank resource (https://www.ukbiobank.ac.uk) under application number 44430.

	\thispagestyle{empty}
	\bibliographystyle{apalike}
	\bibliography{refs-mrsquared}

\begin{thebibliography}{}

\bibitem[Abadie, 2003]{abadie2003semiparametric}
Abadie, A. (2003).
\newblock Semiparametric instrumental variable estimation of treatment response
  models.
\newblock {\em Journal of Econometrics}, 113(2):231--263.

\bibitem[Ackerberg et~al., 2014]{ackerberg2014asymptotic}
Ackerberg, D., Chen, X., Hahn, J., and Liao, Z. (2014).
\newblock Asymptotic efficiency of semiparametric two-step gmm.
\newblock {\em Review of Economic Studies}, 81(3):919--943.

\bibitem[Angrist and Imbens, 1995a]{angrist1995identification}
Angrist, J. and Imbens, G. (1995a).
\newblock Identification and estimation of local average treatment effects.

\bibitem[Angrist and Imbens, 1995b]{angrist1995two}
Angrist, J.~D. and Imbens, G.~W. (1995b).
\newblock Two-stage least squares estimation of average causal effects in
  models with variable treatment intensity.
\newblock {\em Journal of the American statistical Association},
  90(430):431--442.

\bibitem[Angrist et~al., 1996]{angrist1996identification}
Angrist, J.~D., Imbens, G.~W., and Rubin, D.~B. (1996).
\newblock Identification of causal effects using instrumental variables.
\newblock {\em Journal of the American Statistical Association},
  91(434):444--455.

\bibitem[Babino et~al., 2019]{babino2019multiple}
Babino, L., Rotnitzky, A., and Robins, J. (2019).
\newblock Multiple robust estimation of marginal structural mean models for
  unconstrained outcomes.
\newblock {\em Biometrics}, 75(1):90--99.

\bibitem[Bickel et~al., 1993a]{bickel1993efficient}
Bickel, P.~J., Klaassen, C.~A., Bickel, P.~J., Ritov, Y., Klaassen, J.,
  Wellner, J.~A., and Ritov, Y. (1993a).
\newblock {\em Efficient and adaptive estimation for semiparametric models},
  volume~4.
\newblock Johns Hopkins University Press Baltimore.

\bibitem[Bickel et~al., 1993b]{Bickel:1993}
Bickel, P.~J., Klaassen, C.~A., Ritov, Y.~A., Klaassen, J., and Wellner, J.~A.
  (1993b).
\newblock {\em Efficient and adaptive estimation for semiparametric models}.
\newblock Baltimore: Johns Hopkins University Press.

\bibitem[Bowden et~al., 2015]{Bowden:2015aa}
Bowden, J., Davey~Smith, G., and Burgess, S. (2015).
\newblock Mendelian randomization with invalid instruments: effect estimation
  and bias detection through egger regression.
\newblock {\em International Journal of Epidemiology}, 44(2):512--525.

\bibitem[Bowden et~al., 2016a]{Bowden:2016aa}
Bowden, J., Davey~Smith, G., Haycock, P.~C., and Burgess, S. (2016a).
\newblock Consistent estimation in {Mendelian} randomization with some invalid
  instruments using a weighted median estimator.
\newblock {\em Genetic Epidemiology}, 40(4):304--314.

\bibitem[Bowden et~al., 2016b]{bowden2016consistent}
Bowden, J., Davey~Smith, G., Haycock, P.~C., and Burgess, S. (2016b).
\newblock Consistent estimation in {Mendelian} randomization with some invalid
  instruments using a weighted median estimator.
\newblock {\em Genetic Epidemiology}, 40(4):304--314.

\bibitem[Chao and Swanson, 2005]{Chao:2005aa}
Chao, J.~C. and Swanson, N.~R. (2005).
\newblock Consistent estimation with a large number of weak instruments.
\newblock {\em Econometrica}, 73(5):1673--1692.

\bibitem[Chernozhukov et~al., 2018]{Chernozhukov2018ddml}
Chernozhukov, V., Chetverikov, D., Demirer, M., Duflo, E., Hansen, C., Newey,
  W., and Robins, J. (2018).
\newblock {Double/debiased machine learning for treatment and structural
  parameters}.
\newblock {\em The Econometrics Journal}, 21(1):C1--C68.

\bibitem[Chernozhukov et~al., 2022]{chernozhukov2016locally}
Chernozhukov, V., Escanciano, J.~C., Ichimura, H., Newey, W.~K., and Robins,
  J.~M. (2022).
\newblock Locally robust semiparametric estimation.
\newblock {\em Econometrica, forthcoming}.

\bibitem[Davey~Smith and Ebrahim, 2003]{Davey-Smith:2003aa}
Davey~Smith, G. and Ebrahim, S. (2003).
\newblock `{Mendelian} randomization': can genetic epidemiology contribute to
  understanding environmental determinants of disease?
\newblock {\em International Journal of Epidemiology}, 32(1):1--22.

\bibitem[Guo et~al., 2018]{Guo:2018aa}
Guo, Z., Kang, H., Cai, T.~T., and Small, D.~S. (2018).
\newblock Confidence intervals for causal effects with invalid instruments by
  using two-stage hard thresholding with voting.
\newblock {\em Journal of the Royal Statistical Society: Series B (Statistical
  Methodology)}, 80(4):793--815.

\bibitem[Han, 2008]{Han:2008aa}
Han, C. (2008).
\newblock Detecting invalid instruments using {L1-GMM}.
\newblock {\em Economics Letters}, 101(3):285--287.

\bibitem[Hansen, 1982]{Hansen:1982aa}
Hansen, L.~P. (1982).
\newblock Large sample properties of generalized method of moments estimators.
\newblock {\em Econometrica}, 50(4):1029--1054.

\bibitem[Hartwig et~al., 2017]{hartwig2017robust}
Hartwig, F.~P., Davey~Smith, G., and Bowden, J. (2017).
\newblock Robust inference in summary data mendelian randomization via the zero
  modal pleiotropy assumption.
\newblock {\em International journal of epidemiology}, 46(6):1985--1998.

\bibitem[Hausman, 1978]{hausman1978specification}
Hausman, J.~A. (1978).
\newblock Specification tests in econometrics.
\newblock {\em Econometrica}, 46(6):1251--1271.

\bibitem[Heckman, 1997]{heckman1997instrumental}
Heckman, J. (1997).
\newblock Instrumental variables: A study of implicit behavioral assumptions
  used in making program evaluations.
\newblock {\em Journal of Human Resources}, pages 441--462.

\bibitem[Hern{\'a}n and Robins, 2006]{hernan2006instruments}
Hern{\'a}n, M.~A. and Robins, J.~M. (2006).
\newblock Instruments for causal inference: an epidemiologist's dream?
\newblock {\em Epidemiology}, pages 360--372.

\bibitem[Holland, 1988]{holland1988causal}
Holland, P.~W. (1988).
\newblock Causal inference, path analysis and recursive structural equations
  models.
\newblock {\em ETS Research Report Series}, 1988(1):i--50.

\bibitem[Kang et~al., 2020]{kang2020two}
Kang, H., Lee, Y., Cai, T.~T., and Small, D.~S. (2020).
\newblock Two robust tools for inference about causal effects with invalid
  instruments.
\newblock {\em Biometrics}.

\bibitem[Kang et~al., 2016]{Kang:2016aa}
Kang, H., Zhang, A., Cai, T.~T., and Small, D.~S. (2016).
\newblock Instrumental variables estimation with some invalid instruments and
  its application to mendelian randomization.
\newblock {\em Journal of the American Statistical Association},
  111(513):132--144.

\bibitem[Kleiber and Zeileis, 2008]{AppliedEconometricswithR}
Kleiber, C. and Zeileis, A. (2008).
\newblock {\em Applied Econometrics with {R}}.
\newblock Springer-Verlag, New York.
\newblock {ISBN} 978-0-387-77316-2.

\bibitem[Koles{\'a}r et~al., 2015]{kolesar2015identification}
Koles{\'a}r, M., Chetty, R., Friedman, J., Glaeser, E., and Imbens, G.~W.
  (2015).
\newblock Identification and inference with many invalid instruments.
\newblock {\em Journal of Business \& Economic Statistics}, 33(4):474--484.

\bibitem[Lawlor et~al., 2008]{Lawlor:2008aa}
Lawlor, D.~A., Harbord, R.~M., Sterne, J. A.~C., Timpson, N., and Davey~Smith,
  G. (2008).
\newblock Mendelian randomization: Using genes as instruments for making causal
  inferences in epidemiology.
\newblock {\em Statistics in Medicine}, 27(8):1133--1163.

\bibitem[Leeb and P{\"o}tscher, 2008]{leeb2008sparse}
Leeb, H. and P{\"o}tscher, B.~M. (2008).
\newblock Sparse estimators and the oracle property, or the return of hodges'
  estimator.
\newblock {\em Journal of Econometrics}, 142(1):201--211.

\bibitem[Little and Khoury, 2003]{little2003mendelian}
Little, J. and Khoury, M.~J. (2003).
\newblock Mendelian randomisation: a new spin or real progress?
\newblock {\em The Lancet}, 362(9388):930--930.

\bibitem[Liu et~al., 2020a]{liu2020identification}
Liu, L., Miao, W., Sun, B., Robins, J., and Tchetgen, E.~T. (2020a).
\newblock Identification and inference for marginal average treatment effect on
  the treated with an instrumental variable.
\newblock {\em Statistica Sinica}, 30(3):1517.

\bibitem[Liu et~al., 2020b]{liu2020mendelian}
Liu, Z., Ye, T., Sun, B., Schooling, M., and {Tchetgen Tchetgen}, E. (2020b).
\newblock On {Mendelian} randomization mixed-scale treatment effect robust
  identification {(MR MiSTERI)} and estimation for causal inference.
\newblock arXiv:2009.14484.

\bibitem[Locke et~al., 2015]{Locke:2015aa}
Locke, A.~E., Kahali, B., Berndt, S.~I., and et~al. (2015).
\newblock Genetic studies of body mass index yield new insights for obesity
  biology.
\newblock {\em Nature}, 518(7538):197--206.

\bibitem[Molina et~al., 2017]{molina2017multiple}
Molina, J., Rotnitzky, A., Sued, M., and Robins, J. (2017).
\newblock Multiple robustness in factorized likelihood models.
\newblock {\em Biometrika}, 104(3):561--581.

\bibitem[Morrison et~al., 2020]{morrison2020mendelian}
Morrison, J., Knoblauch, N., Marcus, J.~H., Stephens, M., and He, X. (2020).
\newblock Mendelian randomization accounting for correlated and uncorrelated
  pleiotropic effects using genome-wide summary statistics.
\newblock {\em Nature genetics}, 52(7):740--747.

\bibitem[Newey, 1990]{newey1990semiparametric}
Newey, W.~K. (1990).
\newblock Semiparametric efficiency bounds.
\newblock {\em Journal of applied econometrics}, 5(2):99--135.

\bibitem[Newey, 1993]{NEWEY1993419}
Newey, W.~K. (1993).
\newblock Efficient estimation of models with conditional moment restrictions.
\newblock In {\em Econometrics}, volume~11 of {\em Handbook of Statistics},
  pages 419--454. Elsevier.

\bibitem[Newey and McFadden, 1994]{Newey1994_handbook}
Newey, W.~K. and McFadden, D. (1994).
\newblock Chapter 36 large sample estimation and hypothesis testing.
\newblock volume~4 of {\em Handbook of Econometrics}, pages 2111--2245.
  Elsevier.

\bibitem[Newey and Powell, 1999]{newey1999two}
Newey, W.~K. and Powell, J. (1999).
\newblock Two-step estimation, optimal moment conditions, and sample selection
  models.
\newblock {\em Working paper, Department of Economics, Massachusetts Institute
  of Technology}.

\bibitem[Newey and Windmeijer, 2009]{Newey:2009aa}
Newey, W.~K. and Windmeijer, F. (2009).
\newblock Generalized method of moments with many weak moment conditions.
\newblock {\em Econometrica}, 77(3):687--719.

\bibitem[Neyman, 1923]{Neyman:1923a}
Neyman, J. (1923).
\newblock On the application of probability theory to agricultural experiments.
  essay on principles. section 9.
\newblock {\em Statistical Science}, 5(4):465--472. Trans. Dorota M. Dabrowska
  and Terence P. Speed (1990).

\bibitem[Okui et~al., 2012]{okui2012doubly}
Okui, R., Small, D.~S., Tan, Z., and Robins, J.~M. (2012).
\newblock Doubly robust instrumental variable regression.
\newblock {\em Statistica Sinica}, pages 173--205.

\bibitem[Pearl, 2009]{pearl2009causality}
Pearl, J. (2009).
\newblock {\em Causality}.
\newblock Cambridge university press.

\bibitem[Purcell et~al., 2007]{purcell2007plink}
Purcell, S., Neale, B., Todd-Brown, K., Thomas, L., Ferreira, M.~A., Bender,
  D., Maller, J., Sklar, P., De~Bakker, P.~I., Daly, M.~J., et~al. (2007).
\newblock Plink: a tool set for whole-genome association and population-based
  linkage analyses.
\newblock {\em The American Journal of Human Genetics}, 81(3):559--575.

\bibitem[Qi and Chatterjee, 2019]{qi2019mendelian}
Qi, G. and Chatterjee, N. (2019).
\newblock Mendelian randomization analysis using mixture models for robust and
  efficient estimation of causal effects.
\newblock {\em Nature communications}, 10(1):1--10.

\bibitem[Robins, 1994]{robins1994correcting}
Robins, J.~M. (1994).
\newblock Correcting for non-compliance in randomized trials using structural
  nested mean models.
\newblock {\em Communications in Statistics-Theory and Methods},
  23(8):2379--2412.

\bibitem[Robins, 2000]{robins2000robust}
Robins, J.~M. (2000).
\newblock Robust estimation in sequentially ignorable missing data and causal
  inference models.
\newblock In {\em Proceedings of the American Statistical Association}, volume
  1999, pages 6--10. Indianapolis, IN.

\bibitem[Robins et~al., 1992]{robins1992estimating}
Robins, J.~M., Mark, S.~D., and Newey, W.~K. (1992).
\newblock Estimating exposure effects by modelling the expectation of exposure
  conditional on confounders.
\newblock {\em Biometrics}, pages 479--495.

\bibitem[Robins and Ritov, 1997]{robins1997toward}
Robins, J.~M. and Ritov, Y. (1997).
\newblock Toward a curse of dimensionality appropriate (coda) asymptotic theory
  for semi-parametric models.
\newblock {\em Statistics in medicine}, 16(3):285--319.

\bibitem[Robins et~al., 1994]{robins1994estimation}
Robins, J.~M., Rotnitzky, A., and Zhao, L.~P. (1994).
\newblock Estimation of regression coefficients when some regressors are not
  always observed.
\newblock {\em Journal of the American Statistical Association},
  89(427):846--866.

\bibitem[Rosenbaum et~al., 2010]{rosenbaum2010design}
Rosenbaum, P.~R., Rosenbaum, P., and Briskman (2010).
\newblock {\em Design of observational studies}, volume~10.
\newblock Springer.

\bibitem[Rubin, 1974]{Rubin:1974}
Rubin, D.~B. (1974).
\newblock Estimating causal effects of treatments in randomized and
  nonrandomized studies.
\newblock {\em Journal of Educational Psychology}, 6(5):688--701.

\bibitem[Scharfstein et~al., 1999]{10.2307/2669930}
Scharfstein, D.~O., Rotnitzky, A., and Robins, J.~M. (1999).
\newblock Adjusting for nonignorable drop-out using semiparametric nonresponse
  models: Rejoinder.
\newblock {\em Journal of the American Statistical Association},
  94(448):1135--1146.

\bibitem[Small, 2007]{small2007sensitivity}
Small, D.~S. (2007).
\newblock Sensitivity analysis for instrumental variables regression with
  overidentifying restrictions.
\newblock {\em Journal of the American Statistical Association},
  102(479):1049--1058.

\bibitem[Staiger and Stock, 1997]{Staiger:1997aa}
Staiger, D. and Stock, J.~H. (1997).
\newblock Instrumental variables regression with weak instruments.
\newblock {\em Econometrica}, 65(3):557--586.

\bibitem[Stock and Wright, 2000]{stock2000gmm}
Stock, J.~H. and Wright, J.~H. (2000).
\newblock {GMM} with weak identification.
\newblock {\em Econometrica}, 68(5):1055--1096.

\bibitem[Stock et~al., 2002]{Stock:2002aa}
Stock, J.~H., Wright, J.~H., and Yogo, M. (2002).
\newblock A survey of weak instruments and weak identification in generalized
  method of moments.
\newblock {\em Journal of Business \& Economic Statistics}, 20(4):518--529.

\bibitem[Sudlow et~al., 2015]{sudlow2015uk}
Sudlow, C., Gallacher, J., Allen, N., Beral, V., Burton, P., Danesh, J.,
  Downey, P., Elliott, P., Green, J., Landray, M., et~al. (2015).
\newblock {UK Biobank}: an open access resource for identifying the causes of a
  wide range of complex diseases of middle and old age.
\newblock {\em Plos Medicine}, 12(3):e1001779.

\bibitem[Swanson et~al., 2018]{swanson2018partial}
Swanson, S.~A., Hern{\'a}n, M.~A., Miller, M., Robins, J.~M., and Richardson,
  T.~S. (2018).
\newblock Partial identification of the average treatment effect using
  instrumental variables: review of methods for binary instruments, treatments,
  and outcomes.
\newblock {\em Journal of the American Statistical Association},
  113(522):933--947.

\bibitem[Tan, 2010]{tan2010marginal}
Tan, Z. (2010).
\newblock Marginal and nested structural models using instrumental variables.
\newblock {\em Journal of the American Statistical Association},
  105(489):157--169.

\bibitem[Tchetgen~Tchetgen et~al., 2021]{tchetgen2021genius}
Tchetgen~Tchetgen, E., Sun, B., and Walter, S. (2021).
\newblock The {GENIUS} approach to robust {Mendelian} randomization inference.
\newblock {\em Statistical Science}, 36(3):443--464.

\bibitem[Tchetgen~Tchetgen et~al., 2010]{tchetgen2010doubly}
Tchetgen~Tchetgen, E.~J., Robins, J.~M., and Rotnitzky, A. (2010).
\newblock On doubly robust estimation in a semiparametric odds ratio model.
\newblock {\em Biometrika}, 97(1):171--180.

\bibitem[Tchetgen~Tchetgen and Shpitser, 2012]{tchetgen2012semiparametric}
Tchetgen~Tchetgen, E.~J. and Shpitser, I. (2012).
\newblock Semiparametric theory for causal mediation analysis: efficiency
  bounds, multiple robustness, and sensitivity analysis.
\newblock {\em Annals of Statistics}, 40(3):1816.

\bibitem[Tchetgen~Tchetgen et~al., 2015]{tchetgen2015instrumental}
Tchetgen~Tchetgen, E.~J., Walter, S., Vansteelandt, S., Martinussen, T., and
  Glymour, M. (2015).
\newblock Instrumental variable estimation in a survival context.
\newblock {\em Epidemiology (Cambridge, Mass.)}, 26(3):402.

\bibitem[Tibshirani, 2011]{https://doi.org/10.1111/j.1467-9868.2011.00771.x}
Tibshirani, R. (2011).
\newblock Regression shrinkage and selection via the lasso: a retrospective.
\newblock {\em Journal of the Royal Statistical Society: Series B (Statistical
  Methodology)}, 73(3):273--282.

\bibitem[Tsiatis, 2007]{tsiatis2007semiparametric}
Tsiatis, A. (2007).
\newblock {\em Semiparametric theory and missing data}.
\newblock Springer Science \& Business Media.

\bibitem[Tyrrell et~al., 2016]{Tyrrell2016}
Tyrrell, J., Jones, S.~E., Beaumont, R., Astley, C.~M., Lovell, R., Yaghootkar,
  H., Tuke, M., Ruth, K.~S., Freathy, R.~M., Hirschhorn, J.~N., et~al. (2016).
\newblock Height, body mass index, and socioeconomic status: mendelian
  randomisation study in {UK Biobank}.
\newblock {\em BMJ}, 352:i582.

\bibitem[Vansteelandt and Didelez, 2018]{vansteelandt2018improving}
Vansteelandt, S. and Didelez, V. (2018).
\newblock Improving the robustness and efficiency of covariate-adjusted linear
  instrumental variable estimators.
\newblock {\em Scandinavian Journal of Statistics}, 45(4):941--961.

\bibitem[Vansteelandt and Goetghebeur, 2003]{vansteelandt2003causal}
Vansteelandt, S. and Goetghebeur, E. (2003).
\newblock Causal inference with generalized structural mean models.
\newblock {\em Journal of the Royal Statistical Society: Series B (Statistical
  Methodology)}, 65(4):817--835.

\bibitem[Vansteelandt et~al., 2008]{vansteelandt2008multiply}
Vansteelandt, S., VanderWeele, T.~J., Tchetgen~Tchetgen, E.~J., and Robins,
  J.~M. (2008).
\newblock Multiply robust inference for statistical interactions.
\newblock {\em Journal of the American Statistical Association},
  103(484):1693--1704.

\bibitem[Wang and Tchetgen~Tchetgen, 2018]{wang2018bounded}
Wang, L. and Tchetgen~Tchetgen, E. (2018).
\newblock Bounded, efficient and multiply robust estimation of average
  treatment effects using instrumental variables.
\newblock {\em Journal of the Royal Statistical Society: Series B (Statistical
  methodology)}, 80(3):531.

\bibitem[White, 1982]{white1982maximum}
White, H. (1982).
\newblock Maximum likelihood estimation of misspecified models.
\newblock {\em Econometrica: Journal of the econometric society}, pages 1--25.

\bibitem[Windmeijer et~al., 2019]{Windmeijer:2019aa}
Windmeijer, F., Farbmacher, H., Davies, N., and Davey~Smith, G. (2019).
\newblock On the use of the lasso for instrumental variables estimation with
  some invalid instruments.
\newblock {\em Journal of the American Statistical Association},
  114(527):1339--1350.

\bibitem[Wooldridge, 2010]{wooldridge2010econometric}
Wooldridge, J.~M. (2010).
\newblock {\em Econometric analysis of cross section and panel data}.
\newblock MIT press.

\bibitem[Ye et~al., 2021]{ye2021genius}
Ye, T., Liu, Z., Sun, B., and Tchetgen~Tchetgen, E. (2021).
\newblock {GENIUS-MAWII}: For robust {Mendelian} randomization with many weak
  invalid instruments.
\newblock {\em arXiv preprint arXiv:2107.06238}.

\bibitem[Zhao and Yu, 2006]{zhao2006model}
Zhao, P. and Yu, B. (2006).
\newblock On model selection consistency of lasso.
\newblock {\em The Journal of Machine Learning Research}, 7:2541--2563.

\bibitem[Zhao et~al., 2020]{zhao2020statistical}
Zhao, Q., Wang, J., Hemani, G., Bowden, J., and Small, D.~S. (2020).
\newblock Statistical inference in two-sample summary-data mendelian
  randomization using robust adjusted profile score.
\newblock {\em The Annals of Statistics}, 48(3):1742--1769.

\end{thebibliography}
	
\newpage
\begin{center}
{\bf \Large Supplementary Material for ``Semiparametric Efficient G-estimation with Invalid Instrumental Variables"}

\bigskip \bigskip
		
		{\large Baoluo Sun$^1$, Zhonghua Liu$^2$ and Eric Tchetgen Tchetgen$ ^3 $} \\ \bigskip
	{$^1$Department of Statistics and Data Science, National University of Singapore\\
	$^2$Department of Statistics and Actuarial Science, 
	University of Hong Kong\\
	$ ^3 $Department of Statistics and Data Science, The Wharton School of the University of Pennsylvania\\
}
\end{center}
	
\setcounter{equation}{0}
\renewcommand{\theequation}{S\arabic{equation}}
\setcounter{table}{0}
\renewcommand{\thetable}{S\arabic{table}}
\setcounter{figure}{0}
\renewcommand{\thefigure}{S\arabic{figure}}
\setcounter{theorem}{0}
\renewcommand{\thefigure}{S\arabic{theorem}}

\section*{Proof of Proposition 1}

The union causal model $\cup_{\alpha\in\{\ell \subseteq\{1,...,K\}:|\ell|\geq \gamma\}}\mathcal{C}(\alpha)$ implies the union observed data model $\cup_{\alpha\in\{\ell \subseteq\{1,...,K\}:|\ell|\geq \gamma\}}\mathcal{M}(\alpha)$. Furthermore, $\cup_{\alpha\in\{\ell \subseteq\{1,...,K\}:|\ell|>\gamma\}}\mathcal{M}(\alpha)$ implies $\cup_{\alpha\in\{\ell \subseteq\{1,...,K\}:|\ell|=\gamma\}}\mathcal{M}(\alpha)$, so that $\cup_{\alpha\in\{\ell \subseteq\{1,...,K\}:|\ell|\geq \gamma\}}\mathcal{C}(\alpha)$ implies $\cup_{\alpha\in\{\ell \subseteq\{1,...,K\}:|\ell|=\gamma\}}\mathcal{M}(\alpha)$.

\section*{Proof of Theorem 1}

To ease notation, let $\bar{Z}_{k}=Z_k-\mu^{\ast}_k$ denote the demeaned $k$-th instrument. Suppose assumption 2 holds for $K$ candidate instruments. For a fixed value $1\leq\gamma\leq K$, consider the following orthogonal direct sum decomposition,
 \begin{align*}
{\rm I\!H}=&\text{span}\left(\left\{\Pi_{s\in\alpha}\bar{Z}_s:\alpha\in\{\ell \subseteq\{1,...,K\}:1\leq |\ell|\leq K-\gamma  \}\right\}\right)\\
 &\oplus\text{span}\left(\left\{\Pi_{s\in\alpha}\bar{Z}_s:\alpha\in\{\ell \subseteq\{1,...,K\}:K-\gamma+1\leq |\ell|\leq K\}\right\}\right).    
 \end{align*}
Consider any element $h(\boldsymbol{Z})\in \text{span}\left(\left\{\Pi_{s\in\alpha}\bar{Z}_s:\alpha\in\{\ell \subseteq\{1,...,K\}:K-\gamma+1\leq |\ell|\leq K\}\right\}\right)$. For each $\alpha_1\in\left\{\ell \subseteq\{1,...,K\}:K-\gamma+1\leq |\ell|\leq K\}\right\}$ and $\alpha_2\in\left\{\ell \subseteq\{1,...,K\}:|\ell|=\gamma\}\right\}$, $|\alpha_1|>K-\gamma=|\alpha^c_2|$, where $\alpha^c_2=\{k:k\not \in\alpha_2\}$ denotes the complement of set $\alpha_2$. Therefore the set $\alpha_1\setminus \alpha^c_2=\alpha_1\cap \alpha_2$ is non-empty and 
\begin{align*}
E\{ \Pi_{s\in \alpha_1} \bar{Z}_s \rvert \boldsymbol{Z}_{-\alpha_2}\}&=E\{ \Pi_{s\in \alpha_1\cap \alpha_2} \bar{Z}_s \Pi_{s\in \alpha_1\cap \alpha_2^c} \bar{Z}_s  \rvert \boldsymbol{Z}_{-\alpha_2}\}=\Pi_{s\in \alpha_1\cap \alpha_2^c} \bar{Z}_s  E\{ \Pi_{s\in \alpha_1\cap \alpha_2} \bar{Z}_s \}=0, \end{align*} so that $h(\boldsymbol{Z})\in {\rm I\!H}_{\gamma}$. On the other hand, any  element $\upsilon\in{\rm I\!H}_{\gamma}$ can be expressed as $\upsilon=\upsilon_1+\upsilon_2$ with $\upsilon_1\in \text{span}\left(\left\{\Pi_{s\in\alpha}\bar{Z}_s:\alpha\in\{\ell \subseteq\{1,...,K\}:1\leq |\ell|\leq K-\gamma  \}\right\}\right)$ and $$\upsilon_2\in \text{span}\left(\left\{\Pi_{s\in\alpha}\bar{Z}_s:\alpha\in\{\ell \subseteq\{1,...,K\}:K-\gamma+1\leq |\ell|\leq K\}\right\}\right).$$ For each $\alpha_1\in\left\{\ell \subseteq\{1,...,K\}:1\leq |\ell|\leq K-\gamma\}\right\}$, $|\alpha_1|\leq K-\gamma$. Therefore there exists a value of $\alpha_2\in\left\{\ell \subseteq\{1,...,K\}:|\ell|=\gamma\}\right\}$ such that $\alpha_1\subseteq \alpha_2^c$ and
\begin{align*}
E\{ \Pi_{s\in \alpha_1} \bar{Z}_s\rvert \boldsymbol{Z}_{-\alpha_2}\}= \Pi_{s\in \alpha_1} \bar{Z}_s\neq 0,
\end{align*}
almost surely. This implies that $\upsilon_1=0$ and $\upsilon=\upsilon_2$. This proves the first part of the theorem and implies that   the $d_{\gamma}$ (non-redundant) unconditional moment restrictions (6) hold in the union observed data model $\cup_{\alpha\in\{\ell \subseteq\{1,...,K\}:|\ell|= \gamma\}}\mathcal{M}(\alpha)$. A necessary and sufficient rank condition to ensure uniqueness of solution to (6) is given by (7). 

\section*{Proof of Theorem 2}

We follow closely the semiparametric theory of \cite{newey1990semiparametric}, \cite{Bickel:1993} and \cite{tsiatis2007semiparametric} in deriving the influence function of any regular and asymptotically linear estimator $\widehat{\beta}(\gamma)$ of $\beta^{\ast}$ defined by moment restrictions (6). Consider a parametric path $t$ for the joint distribution of $O=(Y,A,\boldsymbol{Z})$ restricted by assumption 2, with joint density function given by $$f_t(y,a,\mathbf{z})=f_t(y,a|\boldsymbol{Z}=\mathbf{z})\Pi_{s=1}^K{\Pr}_t(Z_s=z_s),$$
and $f_0(y,a,\mathbf{z})=f(y,a,\mathbf{z})$. The resulting score function is given by
$$S_t(y,a,\mathbf{z})=S_t(y,a|\mathbf{z})+\sum_{s=1}^KS_t(z_s).$$
The moment restrictions in (6) are equivalent to the requirement that the moment restriction 
\begin{align}
\label{eq:moment}
\Theta^\T E\{\mathbf{D}_{\gamma}(\boldsymbol{Z};\boldsymbol{\mu}^{\ast})(Y-\beta A)\}={0},  
\end{align}
holds for any $d_{\gamma}$-dimensional real vector $\Theta$. If the estimator $\widehat{\beta}(\gamma)$ is regular and asymptotically linear with influence function $\varphi(O)$, then $\partial \beta_t/\partial t\mid_{t=0}=E\{\varphi(O)S(O)\}$ \citep[Theorem 3.2]{tsiatis2007semiparametric}. Differentiating under the integral of (\ref{eq:moment}) yields
\begin{align*}
 \frac{\partial \beta_t}{\partial t}\biggr\rvert_{t=0}=-Q^{-1}\left[\Theta^\T E\{\mathbf{D}_{\gamma}(\boldsymbol{Z};\boldsymbol{\mu}^{\ast})(Y-\beta^{\ast} A)S(O)\}+\frac{\partial \Theta^\T E\{\mathbf{D}_{\gamma}(\boldsymbol{Z};\boldsymbol{\mu}_t)(Y-\beta^{\ast} A)\}}{\partial t}\biggr\rvert_{t=0}\right],
\end{align*}
where $Q=-\Theta^\T E\{\mathbf{D}_{\gamma}(\boldsymbol{Z};\boldsymbol{\mu}^{\ast})A\}$ and $\mu_t=\{E_t(Z_1),...,E_t(Z_K)\}^\T$. The $j$-th entry in $\mathbf{D}_{\gamma}(\boldsymbol{Z};\boldsymbol{\mu}_t)(Y-\beta^{\ast} A)$ is the function $\Pi_{s\in\alpha(j)}\{Z_s-E_t(Z_s)\}(Y-\beta^{\ast} A)$, with corresponding path-wise derivative

\begin{align*}
&\frac{\partial [\Pi_{s\in\alpha(j)}\{Z_s-E_t(Z_s)\}(Y-\beta^{\ast}A)]}{\partial t}\biggr\rvert_{t=0}\\
&=\sum_{1\leq r \leq |\alpha(j)|}\left[\sum_{\alpha\in\{\ell\subseteq\alpha(j):|\ell|=r\}}(-1)^r \frac{\partial E\{\Pi_{s\not\in\alpha,s\in\alpha(j)}Z_sE_t(\Pi_{s\in\alpha}Z_s)(Y-\beta^{\ast} A)\}}{\partial t} \biggr\rvert_{t=0}\right]\\
&=\sum_{1\leq r \leq |\alpha(j)|}\left[\sum_{\alpha\in\{\ell\subseteq\alpha(j):|\ell|=r\}}(-1)^r  E[\Pi_{s\not\in\alpha,s\in\alpha(j)}Z_sE\{\Pi_{s\in\alpha}Z_s S(\boldsymbol{Z}_{\alpha}\mid \boldsymbol{Z}_{-\alpha})\mid \boldsymbol{Z}_{-\alpha}\}(Y-\beta^{\ast} A)]\right]\\
&=\sum_{1\leq r \leq |\alpha(j)|}\left[\sum_{\alpha\in\{\ell\subseteq\alpha(j):|\ell|=r\}}(-1)^r  E\{\Pi_{s\in\alpha(j)}Z_s(Y-\beta^{\ast} A\mid\boldsymbol{Z}_{-\alpha} )S(\boldsymbol{Z}_{\alpha}\mid \boldsymbol{Z}_{-\alpha})\}\right]\\
&=\sum_{1\leq r \leq |\alpha(j)|}\left[\sum_{\alpha\in\{\ell\subseteq\alpha(j):|\ell|=r\}}(-1)^r  E\{\Pi_{s\in\alpha(j)}(Z_s-\mu^{\ast}_s)(Y-\beta^{\ast} A\mid\boldsymbol{Z}_{-\alpha} )S(O)\}\right].
\end{align*}
Therefore
$$ \frac{\partial \beta_t}{\partial t}\biggr\rvert_{t=0}=-Q^{-1}\Theta^\T E[\mathbf{D}_{\gamma}(\boldsymbol{Z};\boldsymbol{\mu}^{\ast})\odot \{(Y-\beta^{\ast} A)+\mathbf{A}_{\gamma}(\boldsymbol{Z};\beta^{\ast})\}S(O)],$$
and the class of influence functions is given by $$\mathcal{F}=\left\{ E\{\Theta^\T\mathbf{D}_{\gamma}(\boldsymbol{Z};\boldsymbol{\mu}^{\ast})A\}^{-1}\Theta^\T \mathbf{G}_{\gamma}(O;\beta^{\ast},\boldsymbol{\mu}^{\ast}):\quad \Theta\in{\rm I\!R}^{d_{\gamma}}, E\{\Theta^\T\mathbf{D}_{\gamma}(\boldsymbol{Z};\boldsymbol{\mu}^{\ast})A\}\neq 0 \right\},$$
which proves the first part of Theorem 2. By Theorem 5.3 of \cite{Newey1994_handbook}, the optimal linear combination in terms of asymptotic variance is indexed by $\Theta_{opt}$ which satisfies the generalized information equality
$$E\{\Theta^\T\mathbf{D}_{\gamma}(\boldsymbol{Z};\boldsymbol{\mu}^{\ast})A\}=E\{\Theta^{\T}\mathbf{G}^{\otimes 2 }_{\gamma}(O;\beta^{\ast},\boldsymbol{\mu}^{\ast})\Theta_{opt}\},\quad \text{ for all }\Theta\in{\rm I\!R}^{d_{\gamma}}.$$
Equivalently, $\Theta^{\T}E\{\mathbf{G}^{\otimes 2 }_{\gamma}(O;\beta^{\ast},\boldsymbol{\mu}^{\ast})\Theta_{opt}-\mathbf{D}_{\gamma}(\boldsymbol{Z};\boldsymbol{\mu}^{\ast})A\}=0 \text{ for all }\Theta\in{\rm I\!R}^{d_{\gamma}}.$ In particular, taking $\Theta=E\{\mathbf{G}^{\otimes 2 }_{\gamma}(O;\beta^{\ast},\boldsymbol{\mu}^{\ast})\Theta_{opt}-\mathbf{D}_{\gamma}(\boldsymbol{Z};\boldsymbol{\mu}^{\ast})A\}$ yields
$$E\{\mathbf{G}^{\otimes 2 }_{\gamma}(O;\beta^{\ast},\boldsymbol{\mu}^{\ast})\Theta_{opt}-\mathbf{D}_{\gamma}(\boldsymbol{Z};\boldsymbol{\mu}^{\ast})A\}^\T E\{\mathbf{G}^{\otimes 2 }_{\gamma}(O;\beta^{\ast},\boldsymbol{\mu}^{\ast})\Theta_{opt}-\mathbf{D}_{\gamma}(\boldsymbol{Z};\boldsymbol{\mu}^{\ast})A\}=0,$$  which implies that $$ \Theta_{opt}=E\{\mathbf{G}^{\otimes 2}_{\gamma}(O;\beta^{\ast},\boldsymbol{\mu}^{\ast})\}^{-1}E\{\mathbf{D}_{\gamma}(\boldsymbol{Z};\boldsymbol{\mu}^{\ast})A\}.$$ The asymptotic variance lower bound for all regular estimators of $\beta^{\ast}$ with influence functions in $\mathcal{F}$ is therefore given by 
\begin{align*}
 \mathcal{V}_{\gamma}&=E\{\Theta^\T_{opt}\mathbf{D}_{\gamma}(\boldsymbol{Z};\boldsymbol{\mu}^{\ast})A\}^{-1}\Theta^\T_{opt} E\{\mathbf{G}^{\otimes  2}_{\gamma}(O;\beta^{\ast},\boldsymbol{\mu}^{\ast})\}\Theta_{opt}E\{\Theta^\T_{opt}\mathbf{D}_{\gamma}(\boldsymbol{Z};\boldsymbol{\mu}^{\ast})A\}^{-1}\\
&=[E\{A\mathbf{D}^\T_{\gamma}(\boldsymbol{Z};\boldsymbol{\mu}^{\ast})\} E\{\mathbf{G}^{\otimes 2}_{\gamma}(O;\beta^{\ast},\boldsymbol{\mu}^{\ast})\}^{-1}E\{\mathbf{D}_{\gamma}(\boldsymbol{Z};\boldsymbol{\mu}^{\ast})A\}]^{-1}.   
\end{align*}

\section*{Proof of Proposition 2}
It follows from the proof of Theorem 1 that the $d_{\gamma}$  unconditional moment restrictions 
\begin{align*}
E\{W(\boldsymbol{Z})\mathbf{D}_{\gamma}(\boldsymbol{Z};\boldsymbol{\mu}^{\ast})(Y-\beta A)\}=\tilde{E}\{\mathbf{D}_{\gamma}(\boldsymbol{Z};\boldsymbol{\mu}^{\ast})(Y-\beta A)\}=\mathbf{0},
\end{align*}
holds in the union observed data model $\cup_{\alpha\in\{\ell \subseteq\{1,...,K\}:|\ell|= \gamma\}}\mathcal{M}(\alpha)$, where the expectation  $\tilde{E}(\cdot)$ is taken with respect to the joint probability mass function $\Pi_{s=1}^K \Pr(Z_k=z_k)$.

\section*{Incorporating Measured Covariates}

Suppose that $(O_1,...,O_n)$ are independent, identically distributed observations of $O=(Y,A,\boldsymbol{Z},\boldsymbol{X})$ from a target population,  where $Y$ is an outcome of interest, $A$ is an exposure, $\boldsymbol{Z}=(Z_1,...,Z_K)$ comprises of  $K$  candidate instruments and $\boldsymbol{X}=(X_1,...,X_p)$ comprises of $p$ measured baseline covariates with support in $\mathcal{X}$. We assume that the following additive linear, constant effects model  holds  with $K$ candidate instruments. \\

{\it Assumption $1^{\prime}$. For two possible values of the exposure $a^{\prime}$, $a$ and a possible value of the instruments $\mathbf{z}\in\mathcal{Z}$ $$Y{(a^{\prime},\mathbf{z})}-Y{(a,\mathbf{0})}=\beta^{\ast} (a^{\prime}-a)+\psi(\mathbf{z},\mathbf{x}),$$
which allows for arbitrary interactions among the direct effects of the instruments within strata defined by the value of baseline covariates $\mathbf{x}\in \mathcal{X}$.}\\

For any index set $\alpha\subseteq \{1,...,K\}$, we say that the candidate instruments   $\boldsymbol{Z}_{\alpha}$ are valid while the remaining instruments  $\boldsymbol{Z}_{-\alpha}$ are invalid if the causal model $\mathcal{C}^{\prime}(\alpha)$ defined by $E\{Y(0,0)|\boldsymbol{Z},\boldsymbol{X}=\mathbf{x}\}=E\{Y(0,0)|\boldsymbol{Z}_{-\alpha},\boldsymbol{X}=\mathbf{x}\}$ and $\psi(\boldsymbol{Z},\boldsymbol{X}=\mathbf{x})=\psi(\boldsymbol{Z}_{-\alpha},\boldsymbol{X}=\mathbf{x})$  holds almost surely for any $\mathbf{x}\in\mathcal{X}$, which implies the observed data model $\mathcal{M}^{\prime}(\alpha)$ defined by the conditional mean independence restriction $$E\{Y-\beta^{\ast}|\boldsymbol{Z},\boldsymbol{X}=\mathbf{x}\}=E\{Y-\beta^{\ast}|\boldsymbol{Z}_{-\alpha},\boldsymbol{X}=\mathbf{x}\},$$
for any $\mathbf{x}\in\mathcal{X}$. We say that { at least } $\gamma$ out of the $K$ candidate instruments are valid if the union causal model 
$\cup_{\alpha\in\{\ell \subseteq\{1,...,K\}:|\ell|\geq \gamma\}}\mathcal{C}^{\prime}(\alpha)$ holds, which implies the union observed data model $\cup_{\alpha\in\{\ell \subseteq\{1,...,K\}:|\ell|= \gamma\}}\mathcal{M}^{\prime}(\alpha)$. In addition, we assume conditional independence of the candidate instruments.\\

{\it Assumption $2^{\prime}$. The conditional distribution of the  candidate instruments  is non-degenerate and factorizes as  $\Pr(\boldsymbol{Z}\leq\mathbf{z}\mid \boldsymbol{X}=\mathbf{x})=\prod_{s=1}^K \Pr(Z_s\leq z_s\mid \boldsymbol{X}=\mathbf{x})$ for any $\mathbf{z}\in\mathcal{Z}$ and $\mathbf{x}\in\mathcal{X}$.}\\

The proof of the following result is similar to the proofs of Theorems 1 and 2.\\

{\it Theorem $3$. Suppose assumption $2^{\prime}$ holds with $K$ candidate instruments and without loss of generality consider the enumeration $\{\alpha(1),...,\alpha(d_{\gamma})\}$ of the elements in $\left\{\ell \subseteq\{1,...,K\}:K-\gamma+1\leq |\ell|\leq K\}\right\}$ in some fixed order. Let $\mathbf{D}_{\gamma}(\boldsymbol{Z},\boldsymbol{X};\boldsymbol{\mu}^{\ast})=[\Pi_{s\in\alpha(1)}\{Z_s-\mu^{\ast}_s(\boldsymbol{X})\},...,\Pi_{s\in\alpha(d_{\gamma})}\{Z_s-\mu^{\ast}_s(\boldsymbol{X})\}]^\T$ where $\mu^{\ast}_s(\boldsymbol{X})=E(Z_s\mid \boldsymbol{X})$ for $s=1,...,K$.  Then the causal parameter of interest $\beta^{\ast}$ is identified in the union observed data model $\cup_{\alpha\in\{\ell \subseteq\{1,...,K\}:|\ell|= \gamma\}}\mathcal{M}^{\prime}(\alpha)$ as the unique solution to the $d_{\gamma}$ conditional moment restrictions
\begin{align}
\label{eq:identx}
E\{\mathbf{D}_{\gamma}(\boldsymbol{Z},\boldsymbol{X};\boldsymbol{\mu}^{\ast})(Y-\beta A)\mid \boldsymbol{X}\}=\mathbf{0},
\end{align}
provided that 
\begin{align}
\label{eq:ident2x}
\biggr|\biggr| \frac{\partial E\{\mathbf{D}_{\gamma}(\boldsymbol{Z},\boldsymbol{X};\boldsymbol{\mu}^{\ast})(Y-\beta A)\mid \boldsymbol{X}=\mathbf{x}\}}{\partial \beta}\biggr|\biggr| _0=||E\{\mathbf{D}_{\gamma}(\boldsymbol{Z},\boldsymbol{X};\boldsymbol{\mu}^{\ast})A\mid \boldsymbol{X}=\mathbf{x}\}||_0> 0,
\end{align}
for at least one value $\mathbf{x}\in\mathcal{X}$. Furthermore,  any regular and asymptotically linear estimator $\widehat{\beta}(\gamma)$ of $\beta^{\ast}$ defined by (\ref{eq:identx}) under assumption $2^\prime$ and condition (\ref{eq:ident2x}) must satisfy the following asymptotic expansion:
\begin{align*}
 {n}^{1/2} \{ \hat{\beta}(\gamma)-\beta^{\ast}\}  = E\{\Theta^\T(\boldsymbol{X})\mathbf{D}_{\gamma}(\boldsymbol{Z},\boldsymbol{X};\boldsymbol{\mu}^{\ast})A\}^{-1} {n}^{1/2}\widehat{E}_n%
\{\Theta^\T(\boldsymbol{X})\mathbf{G}_{\gamma}(O;\beta^{\ast},\boldsymbol{\mu}^{\ast})\}+o_{p}\left(  1\right)  ,   
\end{align*}
for any $d_{\gamma}$-dimensional vector function $\Theta(\boldsymbol{X})$ that satisfies $E\{\Theta^\T(\boldsymbol{X})\mathbf{D}_{\gamma}(\boldsymbol{Z},\boldsymbol{X};\boldsymbol{\mu}^{\ast})A\}\neq 0$, where $\mathbf{G}_{\gamma}(O;\beta,\boldsymbol{\mu})=\mathbf{D}_{\gamma}(\boldsymbol{Z},\boldsymbol{X};\boldsymbol{\mu})\odot \{(Y-\beta A)+\mathbf{A}_{\gamma}(\boldsymbol{Z},\boldsymbol{X};\beta)\}$ and $\mathbf{A}_{\gamma}(\boldsymbol{Z},\boldsymbol{X};\beta)$ is an  $d_{\gamma}$-dimensional augmentation vector with the $j$-th entry equal to
\begin{align*}
\sum_{1\leq r \leq |\alpha(j)|}\left\{\sum_{\alpha\in\{\ell\subseteq\alpha(j):|\ell|=r\}}(-1)^{r} E(Y-\beta A\mid \boldsymbol{Z}_{-\alpha},\boldsymbol{X})\right\}.
\end{align*}
The efficient regular and asymptotically linear estimator estimator admits the above expansion with $\Theta_{opt}(\boldsymbol{X})=E\{\mathbf{G}^{\otimes 2}_{\gamma}(O;\beta^{\ast},\boldsymbol{\mu}^{\ast})\mid \boldsymbol{X}\}^{-1}E\{\mathbf{D}_{\gamma}(\boldsymbol{Z};\boldsymbol{\mu}^{\ast})A\mid \boldsymbol{X}\}$, so that the asymptotic variance lower bound for all regular estimators of $\beta^{\ast}$ is given by
$$\mathcal{V}_{\gamma}=[E\{A\mathbf{D}^\T_{\gamma}(\boldsymbol{Z};\boldsymbol{\mu}^{\ast})\mid \boldsymbol{X}\} E\{\mathbf{G}^{\otimes 2}_{\gamma}(O;\beta^{\ast},\boldsymbol{\mu}^{\ast})\mid \boldsymbol{X}\}^{-1}E\{\mathbf{D}_{\gamma}(\boldsymbol{Z};\boldsymbol{\mu}^{\ast})A\mid \boldsymbol{X}\}]^{-1}.$$
}
\section*{Doubly robust and debiased machine learning estimation}

\subsection*{Doubly robust estimator}
When $\boldsymbol{X}$ is of moderate to high
dimension relative to the sample size, one dimension reduction strategy is to specify working  models $\{\mu_s(\boldsymbol{X};\eta_s): s=1,...,K\}$ and $\mathbf{A}_{\gamma}(\boldsymbol{Z},\boldsymbol{X};\beta,\psi)$ indexed by finite-dimensional parameters $\eta=(\eta^\T_1,...,\eta^\T_K,)^\T$ and $\psi$ respectively, which constrain the nuisance parameter space. A doubly robust estimator $\widehat{\beta}_{dr}(\gamma)$ is obtained as follows.\\

{\it Step 1.} Solve the score function 
$$\widehat{E}_n\left[\left\{{Z_s}/{\mu_s(\boldsymbol{X};\eta_s)}-{(1-Z_s)}/{(1-\mu_s(\boldsymbol{X};\eta_s))}\right\}{\partial\mu_s(\boldsymbol{X};\eta_s) }/{\partial \eta_s}\right]=\mathbf{0},$$
to obtain $\widehat{\eta}_s$ for each $s=1,...,K$. Let $\widehat{\boldsymbol{\mu}}(\boldsymbol{X})=\{\mu_s(\boldsymbol{X};\widehat{\eta}_s):s=1,...,K\}$. \\

{\it Step 2.} For a user specified  positive semi-definite $d_{\gamma}\times d_{\gamma}$ weighting matrix $\Omega$,
\begin{align*}
\widehat{\beta}_{dr}(\gamma)=\arg\min_{\beta} \widehat{E}_n\{\mathbf{G}^{\T}_{\gamma}(O;\beta,\widehat{\boldsymbol{\mu}},\widehat{\psi}(\beta)\}\Omega \widehat{E}_n\{\mathbf{G}_{\gamma}(O;\beta,\widehat{\boldsymbol{\mu}},\widehat{\psi}(\beta))\},
\end{align*}
where $\mathbf{G}_{\gamma}(O;\beta,\widehat{\boldsymbol{\mu}},\widehat{\psi}(\beta))=\mathbf{D}_{\gamma}(\boldsymbol{Z},\boldsymbol{X};\widehat{\boldsymbol{\mu}})\odot \{Y-\beta A +\mathbf{A}_{\gamma}(\boldsymbol{Z},\boldsymbol{X};\beta,\widehat{\psi}(\beta))\}$ and for each fixed value $\beta$, $\widehat{\psi}(\beta)$ solves $\widehat{E}_n[\{Y-\beta A+\mathbf{A}_{\gamma}(\boldsymbol{Z},\boldsymbol{X};\beta,\psi) \}\partial \mathbf{A}_{\gamma}(\boldsymbol{Z},\boldsymbol{X};\beta,\psi)/\partial \psi]=\mathbf{0}$.\\

Asymptotic normality of $\widehat{\beta}_{dr}(\gamma)$ holds by 
 invoking the $n^{-1/2}$ asymptotic expansion for $\widehat{\beta}_{dr}(\gamma)-\beta^{\dag}$ allowing for model
misspecification of the nuisance parameters \citep{white1982maximum}, where $\beta^{\dag}$ denotes the probability limit of $\widehat{\beta}_{dr}(\gamma)$. In addition, let $\boldsymbol{\mu}^{\dag}(\boldsymbol{X})$ and $\mathbf{A}^{\dag}_{\gamma}(\boldsymbol{Z},\boldsymbol{X};\beta^{\dag})$ denote the probability limits of $\widehat{\boldsymbol{\mu}}(\boldsymbol{X})$ and $\mathbf{A}_{\gamma}(\boldsymbol{Z},\boldsymbol{X};\beta^{\dag},\widehat{\psi}(\beta^{\dag}))$ respectively. It follows from Theorem 4 that $\beta^{\dag}=\beta^{\ast}$ if either the model $\{\mu_s(\boldsymbol{X};\eta_s): s=1,...,K\}$ or $\mathbf{A}_{\gamma}(\boldsymbol{Z},\boldsymbol{X};\beta,\psi)$ is correctly specified, but not necessarily both. \\
{
{\it Theorem 4. The $d_{\gamma}$ moment conditions
$${E}[\mathbf{D}_{\gamma}(\boldsymbol{Z},\boldsymbol{X};{\boldsymbol{\mu}}^{\dag})\odot \{Y-\beta^{\ast} A +\mathbf{A}^{\dag}_{\gamma}(\boldsymbol{Z},\boldsymbol{X};\beta^{\ast})\}]=\mathbf{0},$$
hold if at least one of the following holds: (i) ${\boldsymbol{\mu}}^{\dag}={\boldsymbol{\mu}}^{\ast}$ or (ii) $\mathbf{A}^{\dag}_{\gamma}(\boldsymbol{Z},\boldsymbol{X};\beta)=\mathbf{A}_{\gamma}(\boldsymbol{Z},\boldsymbol{X};\beta)$.
}\\

{\it Proof: Let $$\mathbf{A}^{\dag}_{\gamma,j}(\boldsymbol{Z},\boldsymbol{X};\beta)=\sum_{1\leq r \leq |\alpha(j)|}\left\{\sum_{\alpha\in\{\ell\subseteq\alpha(j):|\ell|=r\}}(-1)^{r} E^{\dag}(Y-\beta A\mid \boldsymbol{Z}_{-\alpha},\boldsymbol{X})\right\},$$
denote the $j$-th entry of $\mathbf{A}^{\dag}_{\gamma}(\boldsymbol{Z},\boldsymbol{X};\beta)$. If ${\boldsymbol{\mu}}^{\dag}={\boldsymbol{\mu}}^{\ast}$, then the $j$-th moment condition 
\begin{align*}
E&[\Pi_{s\in\alpha(j)}\{Z_s-\mu^{\ast}_s(\boldsymbol{X})\}\{Y-\beta^{\ast}A+\mathbf{A}^{\dag}_{\gamma,j}(\boldsymbol{Z},\boldsymbol{X};\beta^{\ast})\}] \\
&=E[\Pi_{s\in\alpha(j)}\{Z_s-\mu^{\ast}_s(\boldsymbol{X})\}\{Y-\beta^{\ast}A+\mathbf{A}^{\dag}_{\gamma,j}(\boldsymbol{Z},\boldsymbol{X};\beta^{\ast})-\mathbf{A}^{\dag}_{\gamma,j}(\boldsymbol{Z},\boldsymbol{X};\beta^{\ast})+\mathbf{A}_{\gamma,j}(\boldsymbol{Z},\boldsymbol{X};\beta^{\ast})\}]\\
&=0,    
\end{align*}
where the second equality arise from the fact that for any $1\leq r \leq |\alpha(j)|$ and $\alpha\in\{\ell\subseteq\alpha(j):|\ell|=r\}$, there is at least one instrument indexed by $\zeta\in\alpha\subseteq \alpha(j)$ such that
\begin{align*}
 E&[\{Z_\zeta-\mu^{\ast}_\zeta(\boldsymbol{X})\}\Pi_{s\in\alpha(j),s\neq \zeta}\{Z_s-\mu^{\ast}_s(\boldsymbol{X})\} E^{\dag}(Y-\beta A\mid \boldsymbol{Z}_{-\alpha},\boldsymbol{X})]\\
 &=E[\{Z_\zeta-\mu^{\ast}_\zeta(\boldsymbol{X})\}]E[\Pi_{s\in\alpha(j),s\neq \zeta}\{Z_s-\mu^{\ast}_s(\boldsymbol{X})\} E^{\dag}(Y-\beta A\mid \boldsymbol{Z}_{-\alpha},\boldsymbol{X})]=0,
\end{align*}
and similarly $E[\{Z_\zeta-\mu^{\ast}_\zeta(\boldsymbol{X})\}\Pi_{s\in\alpha(j),s\neq \zeta}\{Z_s-\mu^{\ast}_s(\boldsymbol{X})\} E(Y-\beta A\mid \boldsymbol{Z}_{-\alpha},\boldsymbol{X})]=0$. On the other hand, if 
$\mathbf{A}^{\dag}_{\gamma}(\boldsymbol{Z},\boldsymbol{X};\beta)=\mathbf{A}_{\gamma}(\boldsymbol{Z},\boldsymbol{X};\beta)$ then the $j$-th moment condition 
\begin{align*}
E&[\Pi_{s\in\alpha(j)}\{Z_s-\mu^{\dag}_s(\boldsymbol{X})\}\{Y-\beta^{\ast}A+\mathbf{A}_{\gamma,j}(\boldsymbol{Z},\boldsymbol{X};\beta^{\ast})\}] \\
&=E[\Pi_{s\in\alpha(j)}\{Z_s-\mu^{\ast}_s(\boldsymbol{X})\}\{Y-\beta^{\ast}A+\mathbf{A}_{\gamma,j}(\boldsymbol{Z},\boldsymbol{X};\beta^{\ast})\}] =0,
\end{align*}
where the second equality follows from the fact that for any $\upsilon\subseteq\alpha(j)$ with cardinality $|\upsilon|=\kappa$,
\begin{align*}
E&[\{\Pi_{s\not \in\upsilon,s\in\alpha(j)}Z_s\}\{\Pi_{r\in\upsilon}\mu^{\dag}_r(\boldsymbol{X})\}\{Y-\beta^{\ast}A+\mathbf{A}_{\gamma,j}(\boldsymbol{Z},\boldsymbol{X};\beta^{\ast})\}]\\
&=E[\{\Pi_{s\not \in\upsilon,s\in\alpha(j)}Z_s\}\{\Pi_{r\in\upsilon}\mu^{\dag}_r(\boldsymbol{X})\}E\{Y-\beta^{\ast}A+\mathbf{A}_{\gamma,j}(\boldsymbol{Z},\boldsymbol{X};\beta^{\ast})\mid \boldsymbol{Z}_{-\upsilon},\boldsymbol{X}\}]\\
&=E\Biggr[\{\Pi_{s\not \in\upsilon,s\in\alpha(j)}Z_s\}\{\Pi_{r\in\upsilon}\mu^{\dag}_r(\boldsymbol{X})\}\Biggr\{E(Y-\beta A\mid \boldsymbol{Z}_{-\upsilon},\boldsymbol{X})\\
&\phantom{===}+\sum_{1\leq r \leq |\alpha(j)|}\sum_{\alpha\in\{\ell\subseteq\alpha(j):|\ell|=r\}}(-1)^{r} E(Y-\beta A\mid \boldsymbol{Z}_{-\alpha\cup\upsilon},\boldsymbol{X})\Biggr\}\Biggr]\\
&=E\Biggr[\{\Pi_{s\not \in\upsilon,s\in\alpha(j)}Z_s\}\{\Pi_{r\in\upsilon}\mu^{\dag}_r(\boldsymbol{X})\}\Biggr\{\sum_{t=0}^{\kappa}(-1)^{t}{\kappa \choose t}E(Y-\beta A\mid \boldsymbol{Z}_{-\upsilon},\boldsymbol{X})\\
&\phantom{===}+\sum_{r>\kappa}\sum_{\{\alpha\cup \upsilon\} \in \{\ell\subseteq\alpha(j):|\ell|=r\}}\sum_{t=0}^{\kappa}(-1)^t{\kappa \choose t}E(Y-\beta A\mid \boldsymbol{Z}_{-\alpha\cup\upsilon},\boldsymbol{X})\Biggr\}\Biggr]\\
&=0.
\end{align*}

}
}
\subsection*{Debiased machine learning estimator}

The cross-fitted debiased machine learning approach uses various flexible and data-adaptive methods  to estimate the functions $\{\mu^{\ast}_s(\boldsymbol{X}): s=1,...,K\}$ and $\{\mathbf{m}_y(\boldsymbol{Z},\boldsymbol{X}),\mathbf{m}_a(\boldsymbol{Z},\boldsymbol{X})\}$, where $\mathbf{m}_y(\boldsymbol{Z},\boldsymbol{X})$, $\mathbf{m}_a(\boldsymbol{Z},\boldsymbol{X})$ are $d_{\gamma}$-dimensional vectors with the $j$-th entries equal to
\begin{align*}
\sum_{1\leq r \leq |\alpha(j)|}\left\{\sum_{\alpha\in\{\ell\subseteq\alpha(j):|\ell|=r\}}(-1)^{r} E(Y\mid \boldsymbol{Z}_{-\alpha},\boldsymbol{X})\right\};\\
\sum_{1\leq r \leq |\alpha(j)|}\left\{\sum_{\alpha\in\{\ell\subseteq\alpha(j):|\ell|=r\}}(-1)^{r} E(A\mid \boldsymbol{Z}_{-\alpha},\boldsymbol{X})\right\},
\end{align*}
respectively. Hence, $\mathbf{A}_{\gamma}(\boldsymbol{Z},\boldsymbol{X};\beta)=\mathbf{m}_y(\boldsymbol{Z},\boldsymbol{X})- \beta \mathbf{m}_a(\boldsymbol{Z},\boldsymbol{X})$. Let $(I_j)_{j=1}^{J}$ be a $J$-fold random partition of the observation indices $\{1,2,...,n\}$. A debiased machine learning  estimator $\widehat{\beta}_{dml}(\gamma)$ is obtained as follows. \\

{\it Step 1.} For each $j=1,2,...,J$, obtain $\widehat{\boldsymbol{\mu}}^{(j)}$, $\widehat{\mathbf{m}}^{(j)}_y$ and $\widehat{\mathbf{m}}^{(j)}_a$ by machine learning (e.g. with probability random forest for binary response and regression random forest for continuous response) of the functions $\boldsymbol{\mu}^{\ast}$, $\mathbf{m}_y(\boldsymbol{Z},\boldsymbol{X})$ and $\mathbf{m}_a(\boldsymbol{Z},\boldsymbol{X})$ respectively, using all observations not in $I_j$.\\ 

{\it Step 2.} For a user specified  positive semi-definite $d_{\gamma}\times d_{\gamma}$ weighting matrix $\Omega$, the cross-fitted debiased machine learning  estimator is given by $$\widehat{\beta}_{dml}(\gamma)=\arg\min_{\beta}\widehat{\mathbf{G}}^{\T}_{\gamma}(O;\beta)\Omega\widehat{\mathbf{G}}_{\gamma}(O;\beta),$$
where $\widehat{\mathbf{G}}_{\gamma}(O;\beta)=n^{-1}\sum_{j=1}^J\sum_{i\in I_j} \widehat{\mathbf{G}}^{(j)}_{\gamma}(O_i;\beta)$ and $\widehat{\mathbf{G}}^{(j)}_{\gamma}(O;\beta)=\mathbf{D}_{\gamma}(\boldsymbol{Z},\boldsymbol{X};\widehat{\boldsymbol{\mu}}^{(j)})\odot [Y+\widehat{\mathbf{m}}^{(j)}_y(\boldsymbol{Z},\boldsymbol{X})-\beta\{A+\widehat{\mathbf{m}}^{(j)}_a(\boldsymbol{Z},\boldsymbol{X})\}]$.

Under general regularity conditions established by \citet{Chernozhukov2018ddml,chernozhukov2016locally}, $\widehat{\beta}_{dml}(\gamma)$ is a root-$n$ consistent estimator of $\beta^{\ast}$ if all the nuisance parameters are estimated with mean-squared error rates diminishing faster than $n^{-1/4}$. Such rates are achievable  for many highly data-adaptive machine learning  methods, including LASSO, gradient boosting trees, random forests or ensembles of these methods.

\section*{Additional Monte Carlo results}

\subsection*{Varying Degree of Exposure Interactions}

\noindent It is of interest to investigate the performance of estimators under settings where the exposure model is sparse. In addition to the main effects, we randomly select $\Delta=30$\% or 60\% of both the lower and higher order interactions of order $\geq ||\pi||_0+1$ to have nonzero coefficients in $\tau$. However, this sampling scheme may render the plurality rule invalid even when $\pi=(0,0,0.1,0.2,0.3)^{\T}$, due to potential asymmetric interactions in the exposure model. Therefore, we perform Monte Carlo simulations corresponding to the two cases $\pi=(0,0,0,0.2,0.2)^{\T}$ and $\pi=(0,0,0.2,0.2,0.2)^{\T}$ only, the latter of which represents violation of both the majority and plurality rules. The estimation results are reported in Tables \ref{tab:int} and \ref{tab:int2}. As expected the proposed g-estimator has smaller absolute bias and lower variance as the degree of interactions in the exposure model increases, with empirical coverage close to the nominal level throughout. 
 
\subsection*{Single Index Models}
 
We present further Monte Carlo results for exposures  generated under the class of semiparametric single index models $E(A|Z)=g(\xi^{\T} Z)$ where $g(\cdot)$ is a link function. Specifically, the continuous exposure is  generated without covariates from 
$
A= \exp(\xi^{\T} \boldsymbol{Z})+\varepsilon_2.
$ We also consider the important setting of a binary exposure generated from
$
A= I(-3+\xi^{\T} \boldsymbol{Z}+\varepsilon_2>0),
$
where $I(\cdot)$ is the indicator function. We fix $C=1$, while all other simulation settings are unchanged. Tables \ref{tab:log} and \ref{tab:probit} summarize the estimation results based on 1000 replications. Remarkably, the proposed g-estimator performs well, even though in both settings only main effects are included in the linear predictor function, as exposure interactive effects are induced on the additive scale by the nonlinear link function, rather than specified explicitly {\it a priori}. All other conclusions are qualitatively similar to those drawn when the exposure is generated under the identity link with interactions.

  \begin{table}
\begin{center}
		\caption{Comparison of methods with {continuous exposure generated under the identity link $(C=1.0)$}. The
two rows of results for each estimator correspond to sample sizes of $n = 10000$ and $n = 50000$
respectively.}
		\bigskip
		\label{tab:ident2}
		\begin{tabular}{cccccccccccc}
			\toprule
	    & G-estimator  & Oracle 2SLS & Naive 2SLS &Post-Lasso & TSHT \\
	     \hline\noalign{\smallskip}
&  \multicolumn{5}{c}{Majority rule holds}\\
$|\text{Bias}|$ & 0.004 & 0.000& 0.008 & 0.000 & 0.000 \\
                      & 0.001 & 0.000 & 0.009 & 0.000 & 0.000 \\
$\sqrt{\text{Var}}$ &0.012 & 0.001 & 0.001 & 0.002 & 0.002   \\   
                             &0.006 & 0.001 & 0.001 & 0.001 & 0.001     \\                
$\sqrt{\text{EVar}}$ & 0.011 & 0.001 & 0.001 & 0.002 & 0.001 \\
                               & 0.006 &0.001 & 0.001 & 0.001 & 0.001 \\
Cov95 &0.931 & 0.951 & 0.000 & 0.939 & 0.950 \\
          & 0.941 &0.945 & 0.001 & 0.947 &0.947 \\

&  \multicolumn{5}{c}{Majority rule violated but plurality rule holds}\\
$|\text{Bias}|$ & 0.018 & 0.000& 0.013 & 0.009 & 0.005 \\
                      & 0.010 & 0.000 & 0.013 & 0.007 & 0.000 \\
$\sqrt{\text{Var}}$ & 0.035 & 0.002 & 0.001 & 0.005 & 0.005  \\   
                             &0.026 & 0.001 & 0.001 & 0.004 & 0.002   \\                
$\sqrt{\text{EVar}}$ & 0.041 & 0.002 & 0.001 & 0.005 & 0.005 \\
                               & 0.028 & 0.001 & 0.001 & 0.001 & 0.001 \\
Cov95 & 0.937 & 0.937 & 0.000 & 0.119 & 0.425 \\
          & 0.936 & 0.951 & 0.000 & 0.000 & 0.944 \\
&  \multicolumn{5}{c}{Both majority and plurality rules  violated}\\
$|\text{Bias}|$ & 0.019& 0.000& 0.013 & 0.021 & 0.021  \\
                      & 0.010 & 0.000 & 0.013 & 0.021 & 0.021 \\
$\sqrt{\text{Var}}$ & 0.035 & 0.002 & 0.001 & 0.002 & 0.002   \\   
                             &0.026 & 0.001 & 0.001 & 0.001 & 0.001    \\                
$\sqrt{\text{EVar}}$ & 0.041 & 0.002 & 0.001 & 0.001 & 0.002 \\
                               & 0.028 & 0.001 & 0.001 & 0.001 & 0.001 \\
Cov95 & 0.936 & 0.940 & 0.000 & 0.000 & 0.000 \\
          & 0.936 & 0.950 & 0.000 & 0.000 & 0.000 \\
\hline
		\end{tabular}
   \begin{tablenotes}
      \item {\noindent \small Note: $|\text{Bias}|$ and $\sqrt{\text{Var}}$ are the Monte Carlo absolute bias and standard deviation of the point estimates,  $\sqrt{\text{EVar}}$ is the
square root of the mean of the variance estimates and Cov95 is the coverage proportion of
the 95\% confidence intervals, based on 1000 repeated simulations. Zeros denote values smaller than $.0005$.}
    \end{tablenotes}

	\end{center}
  \end{table}

  \begin{table}
\begin{center}
		\caption{ Comparison of methods with {continuous exposure generated under the identity link $(C=0.6)$}, and varying degree of interactive effects. The
two rows of results for each estimator correspond to sample sizes of $n = 10000$ and $n = 50000$
respectively.}
		\bigskip
		\label{tab:int}
		\begin{tabular}{cccccccccccc}
			\toprule
	    & G-estimator & Oracle 2SLS & Naive 2SLS &Post-Lasso & TSHT \\
	     \hline\noalign{\smallskip}
&  \multicolumn{5}{c}{Majority rule holds, $\Delta=30$\%}\\
$|\text{Bias}|$ &$0.029$ & 0.000 & $0.039$ & $0.002$ & $0.002$ \\ 
&$0.009$ & 0.000 & $0.039$ &0.000 &  0.000 \\ 
$\sqrt{\text{Var}}$ &$0.056$ & $0.006$ & $0.007$ & $0.009$ & $0.010$ \\ 
&$0.030$ & $0.003$ & $0.006$ & $0.003$ & $0.003$ \\ 
$\sqrt{\text{EVar}}$ &$0.057$ & $0.007$ & $0.005$ & $0.007$ & $0.007$ \\ 
&$0.031$ & $0.003$ & $0.002$ & $0.003$ & $0.003$ \\ 
Cov95 & $0.919$ & $0.961$ & 0.000 & $0.937$ & $0.919$ \\ 
&$0.940$ & $0.941$ & 0.000 & $0.941$ & $0.943$ \\ 

&  \multicolumn{5}{c}{Majority rule holds, $\Delta=60$\%}\\
$|\text{Bias}|$ &$0.012$ & 0.000 & $0.022$ & 0.000 & 0.000 \\ 
&$0.003$ & 0.000 & $0.022$ & 0.000& 0.000 \\ 
$\sqrt{\text{Var}}$ &$0.032$ & $0.004$ & $0.003$ & $0.004$ & $0.005$ \\ 
&$0.016$ & $0.002$ & $0.002$ & $0.002$ & $0.003$ \\ 
$\sqrt{\text{EVar}}$ &$0.031$ & $0.004$ & $0.003$ & $0.004$ & $0.004$ \\ 
&$0.016$ & $0.002$ & $0.001$ & $0.002$ & $0.002$ \\ 
Cov95 &$0.927$ & $0.953$ & 0.000 & $0.954$ & $0.951$ \\ 
&$0.948$ & $0.950$ & 0.000 & $0.945$ & $0.944$ \\ 
&  \multicolumn{5}{c}{Both majority and plurality rules  violated, $\Delta=30$\%}\\
$|\text{Bias}|$ &$0.098$ & 0.000 & $0.059$ & $0.093$ & $0.085$ \\ 
&$0.070$ & 0.000 & $0.059$ & $0.094$ & $0.054$ \\ 
$\sqrt{\text{Var}}$ &$0.143$ & $0.009$ & $0.007$ & $0.016$ & $0.029$ \\ 
&$0.125$ & $0.004$ & $0.006$ & $0.011$ & $0.037$ \\ 
$\sqrt{\text{EVar}}$ &$0.166$ & $0.009$ & $0.005$ & $0.007$ & $0.008$ \\ 
&$0.137$ & $0.004$ & $0.002$ & $0.003$ & $0.003$ \\ 
Cov95 &$0.923$ & $0.943$ & 0.000 & 0.000 & $0.024$ \\ 
&$0.922$ & $0.941$ & 0.000 & 0.000 & $0.205$ \\  
          &  \multicolumn{5}{c}{Both  majority and plurality rules violated, $\Delta=60$\%}\\
$|\text{Bias}|$ &$0.054$ & 0.000 & $0.034$ & $0.055$ & $0.054$ \\ 
&$0.030$ & 0.000 & $0.034$ & $0.055$ & $0.042$ \\ 
$\sqrt{\text{Var}}$ &$0.091$ & $0.005$ & $0.004$ & $0.006$ & $0.009$ \\ 
&$0.072$ & $0.002$ & $0.002$ & $0.004$ & $0.017$ \\ 
$\sqrt{\text{EVar}}$ &$0.106$ & $0.005$ & $0.003$ & $0.004$ & $0.004$ \\ 
&$0.077$ & $0.002$ & $0.001$ & $0.002$ & $0.002$ \\ 
Cov95 &$0.936$ & $0.946$ & 0.000 & 0.000 & 0.000 \\ 
&$0.925$ & $0.959$ & 0.000 & 0.000 & $0.043$ \\ 
\hline
		\end{tabular}
   \begin{tablenotes}
  \item {\noindent \small Note: See the footnote of Table \ref{tab:ident2}.}
    \end{tablenotes}
    
	\end{center}
  \end{table}

  \begin{table}
	\begin{center}
		\caption{Comparison of methods with {continuous exposure generated under the identity link $(C=10.0)$}, and varying degree of interactive effects. The
two rows of results for each estimator correspond to sample sizes of $n = 10000$ and $n = 50000$
respectively.}
		\bigskip
		\label{tab:int2}
		\begin{tabular}{cccccccccccc}
			\toprule
	    & G-estimator & Oracle 2SLS & Naive 2SLS &Post-Lasso & TSHT \\
	     \hline\noalign{\smallskip}
&  \multicolumn{5}{c}{Majority rule holds, $\Delta=30$\%}\\
$|\text{Bias}|$ &$0.015$ & 0.000 & $0.024$ & $0.001$ & $0.001$ \\ 
&$0.005$ & 0.000 & $0.024$ & 0.000&0.000 \\ 
$\sqrt{\text{Var}}$ &$0.034$ & $0.004$ & $0.004$ & $0.006$ & $0.006$ \\ 
&$0.018$ & $0.002$ & $0.003$ & $0.002$ & $0.002$ \\ 
$\sqrt{\text{EVar}}$ &$0.035$ & $0.004$ & $0.003$ & $0.004$ & $0.004$ \\ 
&$0.019$ & $0.002$ & $0.001$ & $0.002$ & $0.002$ \\ 
Cov95 &$0.932$ & $0.961$ & 0.000 & $0.937$ & $0.919$ \\ 
&$0.942$ & $0.941$ & 0.000 & $0.941$ & $0.942$ \\ 

&  \multicolumn{5}{c}{Majority rule holds, $\Delta=60$\%}\\
$|\text{Bias}|$ &$0.006$ &  0.000 & $0.013$ &  0.000 &  0.000 \\ 
&$0.002$ &  0.000 & $0.013$ &  0.000 &  0.000 \\ 
$\sqrt{\text{Var}}$ &$0.020$ & $0.002$ & $0.002$ & $0.002$ & $0.003$ \\ 
&$0.009$ & $0.001$ & $0.001$ & $0.001$ & $0.002$ \\ 
$\sqrt{\text{EVar}}$ &$0.019$ & $0.002$ & $0.002$ & $0.002$ & $0.002$ \\ 
&$0.009$ & $0.001$ & $0.001$ & $0.001$ & $0.001$ \\ 
Cov95 &$0.932$ & $0.953$ &  0.000 & $0.954$ & $0.950$ \\ 
&$0.950$ & $0.950$ &  0.000 & $0.945$ & $0.944$ \\  
&  \multicolumn{5}{c}{Both majority and plurality rules  violated, $\Delta=30$\%}\\
$|\text{Bias}|$ &$0.056$ &  0.000 & $0.036$ & $0.056$ & $0.051$ \\ 
&$0.040$ &  0.000 & $0.035$ & $0.056$ & $0.033$ \\ 
$\sqrt{\text{Var}}$ &$0.088$ & $0.005$ & $0.004$ & $0.010$ & $0.017$ \\ 
&$0.077$ & $0.002$ & $0.003$ & $0.007$ & $0.022$ \\ 
$\sqrt{\text{EVar}}$ &$0.104$ & $0.005$ & $0.003$ & $0.004$ & $0.005$ \\ 
&$0.084$ & $0.002$ & $0.001$ & $0.002$ & $0.002$ \\ 
Cov95 &$0.935$ & $0.944$ &  0.000 &  0.000 & $0.023$ \\ 
&$0.931$ & $0.943$ &  0.000 &  0.000 & $0.205$ \\ 
          &  \multicolumn{5}{c}{Both  majority and plurality rules violated, $\Delta=60$\%}\\
$|\text{Bias}|$ &$0.030$ &  0.000 & $0.020$ & $0.033$ & $0.033$ \\ 
&$0.017$ &  0.000 & $0.020$ & $0.033$ & $0.025$ \\ 
$\sqrt{\text{Var}}$ &$0.055$ & $0.003$ & $0.002$ & $0.003$ & $0.006$ \\ 
&$0.044$ & $0.001$ & $0.001$ & $0.002$ & $0.010$ \\ 
$\sqrt{\text{EVar}}$ &$0.064$ & $0.003$ & $0.002$ & $0.002$ & $0.003$ \\ 
&$0.047$ & $0.001$ & $0.001$ & $0.001$ & $0.001$ \\ 
Cov95 &$0.944$ & $0.946$ &  0.000&  0.000 &  0.000 \\ 
&$0.929$ & $0.959$ &  0.000 &  0.000 & $0.041$ \\ 
\hline
		\end{tabular}
   \begin{tablenotes}
  \item {\noindent\small Note: See the footnote of Table \ref{tab:ident2}.}
    \end{tablenotes}
	\end{center}
  \end{table}

  \begin{table}
\begin{center}
		\caption{Comparison of methods with {continuous exposure generated under the log link}. The
two rows of results for each estimator correspond to sample sizes of $n = 10000$ and $n = 50000$
respectively.}
		\bigskip
		\label{tab:log}
		\begin{tabular}{cccccccccccc}
			\toprule
	    & G-estimator  & Oracle 2SLS & Naive 2SLS &Post-Lasso & TSHT \\
	    \hline\noalign{\smallskip}
&  \multicolumn{5}{c}{Majority rule holds}\\
$|\text{Bias}|$ &$0.000$ & $0.000$ & $0.002$ & $0.000$& $0.000$ \\ 
&$0.000$ & $0.000$ & $0.002$ & $0.000$&$0.000$ \\ 
$\sqrt{\text{Var}}$ &$0.001$ & $0.000$ & $0.000$& $0.000$ & $0.000$ \\ 
&$0.001$ & $0.000$ & $0.000$ &$0.000$ &$0.000$ \\ 
$\sqrt{\text{EVar}}$ &$0.001$ & $0.000$ & $0.000$& $0.000$ & $0.000$ \\ 
&$0.001$ & $0.000$& $0.000$ & $0.000$ & $0.000$ \\ 
Cov95 &$0.939$ & $0.951$ & $0.000$ & $0.938$ & $0.950$ \\ 
&$0.948$ & $0.945$ & $0.000$ & $0.947$ & $0.946$ \\ 
&  \multicolumn{5}{c}{Majority rule violated but plurality rule holds}\\
$|\text{Bias}|$ &$0.002$ & $0.000$ & $0.003$ & $0.002$ & $0.001$ \\ 
&$0.000$ & $0.000$ & $0.003$ & $0.001$ & $0.000$ \\ 
$\sqrt{\text{Var}}$ &$0.006$ & $0.000$ & $0.000$ & $0.001$ & $0.001$ \\ 
&$0.003$ & $0.000$ & $0.000$ & $0.001$ & $0.000$ \\ 
$\sqrt{\text{EVar}}$ &$0.006$ & $0.000$ &$0.000$& $0.000$& $0.000$ \\ 
&$0.003$ & $0.000$ &$0.000$&$0.000$ &$0.000$\\ 
Cov95 &$0.944$ & $0.939$ &$0.000$ & $0.118$ & $0.423$ \\ 
&$0.956$ & $0.951$ & $0.000$ & $0.000$& $0.944$ \\ 
&  \multicolumn{5}{c}{Both majority and plurality rules  violated}\\
$|\text{Bias}|$ &$0.002$ & $0.000$ & $0.003$ & $0.004$ & $0.004$ \\ 
&$0.000$ &$0.000$& $0.003$ & $0.004$ & $0.004$ \\ 
$\sqrt{\text{Var}}$ &$0.006$ & $0.000$ & $0.000$ & $0.000$ &$0.000$ \\ 
&$0.003$ & $0.000$ & $0.000$ & $0.000$ & $0.000$ \\ 
$\sqrt{\text{EVar}}$ &$0.006$ & $0.000$ & $0.000$ & $0.000$ & $0.000$\\ 
&$0.003$ & $0.000$ & $0.000$ & $0.000$& $0.000$ \\ 
Cov95 &$0.940$ & $0.939$ &$0.000$ &$0.000$ & $0.000$ \\ 
&$0.956$ & $0.950$ & $0.000$ &$0.000$ & $0.000$  \\ 
\hline
		\end{tabular}
   \begin{tablenotes}
      \item  {\noindent \small Note: See the footnote of Table \ref{tab:ident2}.}
    \end{tablenotes}
	\end{center}
  \end{table}

  \begin{table}
\begin{center}
		\caption{Comparison of methods with {binary exposure generated under the probit link}. The
two rows of results for each estimator correspond to sample sizes of $n = 10000$ and $n = 50000$
respectively.}
		\bigskip
		\label{tab:probit}
		\begin{tabular}{cccccccccccc}
			\toprule
	    & G-estimator  & Oracle 2SLS & Naive 2SLS &Post-Lasso & TSHT \\
	     \hline\noalign{\smallskip}
&  \multicolumn{5}{c}{Majority rule holds}\\
$|\text{Bias}|$ &$0.065$ & $0.002$ & $0.300$ & $0.002$ & $0.000$ \\ 
&$0.015$ & $0.002$ & $0.302$ & $0.002$ & $0.004$ \\ 
$\sqrt{\text{Var}}$ &$0.187$ & $0.051$ & $0.039$ & $0.055$ & $0.059$ \\ 
&$0.090$ & $0.023$ & $0.018$ & $0.023$ & $0.032$ \\ 
$\sqrt{\text{EVar}}$ &$0.185$ & $0.051$ & $0.039$ & $0.051$ & $0.052$ \\ 
&$0.091$ & $0.023$ & $0.017$ & $0.022$ & $0.023$ \\ 
Cov95 &$0.928$ & $0.952$ & $0.000$ & $0.941$ & $0.948$ \\ 
&$0.954$ & $0.947$ & $0.000$ & $0.948$ & $0.946$ \\ 
&  \multicolumn{5}{c}{Majority rule violated but plurality rule holds}\\
$|\text{Bias}|$ &$0.377$ & $0.003$ & $0.450$ & $0.302$ & $0.203$ \\ 
&$0.196$ & $0.003$ & $0.453$ & $0.252$ & $0.008$ \\ 
$\sqrt{\text{Var}}$ &$0.704$ & $0.062$ & $0.040$ & $0.196$ & $0.195$ \\ 
&$0.535$ & $0.028$ & $0.018$ & $0.147$ & $0.053$ \\ 
$\sqrt{\text{EVar}}$ &$1.077$ & $0.062$ & $0.039$ & $0.056$ & $0.062$ \\ 
&$0.561$ & $0.028$ & $0.017$ & $0.026$ & $0.027$ \\ 
Cov95 &$0.932$ & $0.939$ & $0.000$ & $0.110$ & $0.396$ \\ 
&$0.925$ & $0.951$ & $0.000$ & $0.000$ & $0.945$ \\ 
&  \multicolumn{5}{c}{Both majority and plurality rules  violated}\\
$|\text{Bias}|$ &$0.378$ & $0.003$ & $0.450$ & $0.749$ & $0.751$ \\ 
&$0.196$ & $0.003$ & $0.453$ & $0.753$ & $0.752$ \\ 
$\sqrt{\text{Var}}$ &$0.704$ & $0.062$ & $0.039$ & $0.055$ & $0.055$ \\ 
&$0.536$ & $0.028$ & $0.018$ & $0.023$ & $0.030$ \\ 
$\sqrt{\text{EVar}}$ &$1.085$ & $0.062$ & $0.039$ & $0.051$ & $0.051$ \\ 
&$0.561$ & $0.028$ & $0.017$ & $0.022$ & $0.023$ \\ 
Cov95 &$0.933$ & $0.939$ & $0.000$ & $0.000$ &$0.000$ \\ 
&$0.924$ & $0.951$ & $0.000$ & $0.000$ & $0.000$ \\  
\hline
		\end{tabular}
   \begin{tablenotes}
      \item  {\noindent \small Note: See the footnote of Table \ref{tab:ident2}.}
    \end{tablenotes}
	\end{center}
  \end{table}

	
\end{document}